\documentclass[12pt,eqsecnum]{article}

\usepackage[dvips]{graphicx}
\usepackage{amssymb}
\usepackage{amsmath}
\usepackage{epstopdf}
\makeatletter

\@addtoreset{equation}{section}
\makeatletter

\newcommand{\ma}[1]{\mbox{$\mathcal{#1}$}}

\newcommand{\D}{{\rm d}}

\newcommand{\dalm}{\kern1pt\vbox{\hrule height 0.9pt\hbox{\vrule width
0.9pt\hskip 2.5pt\vbox{\vskip 5.5pt}\hskip 3pt\vrule width 0.3pt}\hrule height
0.3pt}\kern1pt}

\def\b2hat{ {\hat b}_2 }

\def\HRZ{\eta(R,Z)}

\def\Hb{w^{(m)}_{(I)}(b)}
\def\hw{s(\nu)}

\newcommand{\Ri}{{\cal R}}

\def\Ri{{\cal R}}

\newcommand{\ba}{\begin{array}}
\newcommand{\ea}{\end{array}}
\newcommand{\be}{\begin{equation}}
\newcommand{\ee}{\end{equation}}
\newcommand{\bea}{\begin{eqnarray}}
\newcommand{\eea}{\end{eqnarray}}

\newcommand{\Ialpha}{{\beta^{(I)}}}
\newcommand{\IW}{W_{(I)}}

\begin{document}

\begin{titlepage}
\vfill
\begin{flushright}
\today
\end{flushright}

\vfill
\begin{center}
\baselineskip=16pt
{\Large\bf 
New 2D dilaton gravity for nonsingular black holes\\
}
\vskip 0.5cm
{\large {\sl }}
\vskip 10.mm
{\bf Gabor Kunstatter${}^{a}$, Hideki Maeda${}^{b}$, and Tim Taves${}^{c}$} \\

\vskip 1cm
{
	${}^a$ Department of Physics, University of Winnipeg and Winnipeg Institute for Theoretical Physics, Winnipeg, Manitoba, Canada R3B 2E9\\
	${}^b$ Department of Electronics and Information Engineering, Hokkai-Gakuen University, Sapporo 062-8605, Japan\\
	${}^c$ Centro de Estudios Cient\'{\i}ficos (CECs), Casilla 1469, Valdivia, Chile. \\
	\texttt{g.kunstatter-at-uwinnipeg.ca, h-maeda-at-hgu.jp, timtaves-at-gmail.com}

     }
\vspace{6pt}
\today
\end{center}
\vskip 0.2in
\par
\begin{center}
{\bf Abstract}
\end{center}
\begin{quote}
We construct a two-dimensional action that is an extension of spherically symmetric Einstein-Lanczos-Lovelock gravity. 
The action contains arbitrary functions of the areal radius and the norm squared of its gradient, but the field equations are second order and obey Birkhoff's theorem. 
In complete analogy with spherically symmetric Einstein-Lanczos-Lovelock gravity, the field equations admit the generalized Misner-Sharp mass as the first integral that  determines the form of the vacuum solution. 
The arbitrary functions in the action allow for  vacuum solutions that describe a larger class of interesting nonsingular black-hole spacetimes than previously available. 
\vskip 2.mm
\end{quote}
\end{titlepage}




\tableofcontents

\newpage

\section{Introduction}
General relativity has to date passed all experimental and
observational tests. However, the singularity theorems in classical general
relativity imply that the appearance of singularities is generically an inevitable result of
gravitational collapse of massive stars. This powerful result leads to the conclusion that there must be spacetime regions in our universe where the curvature is so large that general relativity is no longer reliable. In such highly curved spacetime regions near the Big-Bang or deep inside black holes, quantum gravitational effects must play a large role in the description of spacetime. While it is strongly believed that the singularities of classical general relativity are ultimately  cured by quantum gravity, a complete quantum theory of gravity is not yet at hand.

This situation is somewhat analogous to that of the early 20th century when Rutherford's classical model of atoms faced the problem of the electromagnetic radiation instability. We now know that this serious problem in classical physics can only be solved by invoking quantum mechanics. It is nonetheless true that in the semi-classical equations of motion, quantum effects can be incorporated as an effective repulsive force that balances the attractive electromagnetic force between proton and electron. A detailed analysis of effective, semi-classical equations for the hydrogen atom based on the method of moments~\cite{Bojowald2006} can be found in~\cite{Chacon-Acosta2012}. Another suggestive result~\cite{Das2013} shows that quantum corrections to the Raychaudhuri equation prevent focusing of geodesics and the formation of conjugate points. 
Based on this potential resolution of the instability problem in terms of modified classical equations of motion, it is reasonable to expect that the singularity problem in general relativity can also be addressed by considering suitable modified, semi-classical theories of gravity.

The spherically symmetric Schwarzschild black hole is a typical vacuum solution containing a spacelike curvature singularity. Generally, there are two possible global structures for nonsingular
black holes that are obtained by modifying the Schwarzschild black hole. In the first class, the spacelike singularity  is replaced by a regular Big-Bounce so that the spacetime in the interior is extended to a cosmological spacetime beyond the bounce. This class of nonsingular black holes has been obtained as exact solutions to a modified theory based on the polymer quantization of gravity~\cite{Peltola2009a,Peltola2009b}.

By contrast, nonsingular black holes in the second class contain a regular de~Sitter core and have  global structures similar to that of the Reissner-Nordstr\"om black hole, but with a regular center.  Sakharov~\cite{Sakharov1966} was first to suggest that black holes might be nonsingular with a de~Sitter core, while Bardeen~\cite{Bardeen1968} presented an explicit metric for such a spacetime, noting that it necessarily requires a topology change at the center. Israel and Poisson~\cite{Poisson1988} derived a similar nonsingular metric from considerations of semi-classical quantum gravity, while the stability of black holes with de~Sitter cores has been extensively analyzed by Dymnikova and Galaktionov~\cite{Dymnikova2005}. Such nonsingular black holes have been realized as exact solutions in general relativity with  particular matter
fields, such as a nonlinear electromagnetic field~\cite{Ayon-Beato1999,Garcia99,Ayon-Beato1999a,Ayon-Beato1999a,Ayon-Beato2000,Ayon-Beato2005}.

At present, it is not exactly known what types of modified gravity can be realized as effective theories of quantum gravity. Under these circumstances having at one's disposal a large class of modified gravity theories
admitting nonsingular black holes provides a firm ground for future studies of the singularity problem and its related puzzles such as the information loss problem.

In the present paper, we focus on two-dimensional (2D) dilaton gravity because it provides  an effective theory for spherically symmetric spacetimes in higher dimensions. Indeed, it has been shown that particular types of 2D dilaton gravity~\cite{Grumiller2002} admit nonsingular black holes as exact solutions~\cite{Ziprick2010, Ziprick2009, Taves2014,Grumiller2003, Grumiller2004}.
The purpose of this paper is to present a new and considerably larger class of 2D dilaton
gravity theories that satisfy Birkhoff's theorem. We show that there exist members of this class of theories that admit nonsingular black holes with a de~Sitter core as unique vacuum solutions and with maximal curvature bounded below for arbitrarily large mass\footnote{GK is grateful to Valeri Frolov for impressing on him the importance of this latter criterion.}. This latter feature was, to the best of our knowledge, not possible in the context of ordinary 2D dilaton gravity without the addition of matter.


The paper is organized as follows: In the next section we review the forms of spherically symmetric Einstein gravity and Einstein-Lanczos-Lovelock (ELL) gravity that we wish to consider. 
Section~\ref{sec2} presents our generalization of spherically symmetric ELL gravity, derives the mass functions and proves Birkhoff's theorem. The Hamiltonian analysis in our new theories is performed in section~\ref{sec:hamilton}.
Section~\ref{sec5} shows how to derive specific nonsingular black holes of physical interest, and also defines a subclass of theories, dubbed {\it designer Lovelock gravity} that are more closely connected to the original ELL gravity\footnote{Note that there is no relationship between the class of theories we consider and designer gravity, as first introduced in~\cite{Hertog2005}. In the latter work, one is concerned with the  defining gravity theories that admit arbitrary boundary conditions generally in the context of AdS/CFT duality.}.
Finally we close with a summary and  prospects for future work. 
Details of lengthy calculations in the Hamiltonian analysis are presented in Appendix~\ref{appendix}.
We adopt units such that $c=\hbar=1$.

\section{2D effective actions for spherically symmetric systems}
\label{sec:2D}
In this section, we review the effective 2D actions for spherically symmetric general relativity and ELL gravity.
As shown in the following section, our new 2D gravity generalizes the latter just as 2D dilaton gravity is a generalization of the former.

\subsection{Einstein gravity}
The Einstein-Hilbert action for general relativity in arbitrary $n$ dimensions is given by 
\be
\label{eq:Einstein}
I_{{\rm EH}} = \frac{1}{16\pi G_{(n)}}\int \D^n{x} \sqrt{-{g}} {\cal R}[{g}],
\ee
where $G_{(n)}$ is the higher dimensional gravitational constant and ${\Ri}[{g}]$ is the Ricci scalar calculated using the $n$-dimensional metric, $g$.  
The most general metric for $n$-dimensional spherically symmetric spacetimes is given by
\begin{align}
\D s_{(n)}^2 =& g_{\mu\nu}({x})\D {x}^\mu \D {x}^\nu \nonumber \\
=& {\bar g}_{AB}({y})\D {y}^A \D {y}^B + R({y})^2 \D \Omega_{(n-2)}^2, \label{eq:higherDg}
\end{align}
where ${\bar g}_{AB}({y})~(A,B=0,1)$ is the general 2D Lorentzian metric, $\D \Omega_{(n-2)}^2$ is the line-element on the unit $(n-2)$-sphere, and $R$ is the areal radius.
After imposing spherical symmetry and integrating out the angular variables the action (\ref{eq:Einstein}) takes the form~\cite{Maeda2008}
\begin{align}
I_{(2)}&=\frac{1}{l^{n-2}}\int \D^2{y} \sqrt{-{\bar g}}\Big\{ R^{n-2}\Ri[{\bar g}]  + (n-2)(n-3)R^{n-4}(D R)^2  + (n-2)(n-3)R^{n-4}\Big\},
\label{eq:dim_reduced_action}
\end{align}
where $(DR)^2:=(D_A R)(D^AR)$ in which $D_A$ is the 2D covariant derivative, and $\Ri[{\bar g}]$ is the Ricci scalar of ${\bar g}$, the 2D Lorentzian part of the higher dimensional metric.
We have defined a length parameter $l$ proportional to the Planck length:
\be
l^{n-2}:= \frac{16\pi G_{(n)}}{{\cal A}_{(n-2)}},
\ee
where ${\cal A}_{(n-2)}$ is the invariant volume of a unit $(n-2)$-sphere.   The variation of the action \eqref{eq:dim_reduced_action} will give the same equations of motion as varying \eqref{eq:Einstein} and then implementing the spherical symmetry.  This is the case whenever the symmetry group is a compact lie group, as is the case in this paper\cite{Palais1979, Fels2002, Deser2003a}.

This system admits the Misner-Sharp quasi-local mass~\cite{Misner-Sharp}:
\bea
{\cal M}:=\frac{(n-2) R^{n-3}}{l^{n-2}}[1-(DR)^2],
\eea
which satisfies $D_A{\cal M}=0$ and hence ${\cal M}$ is constant in vacuum.
By Birkhoff's theorem, the unique vacuum solution is the well-known Schwarzschild-Tangherlini solution:
\bea
\D s_{(n)}^2 &=& -\left(1-\frac{l^{n-2}M}{(n-2) R^{n-3}}\right)\D t^2 + \left(1-\frac{l^{n-2}M}{(n-2) R^{n-3}}\right)^{-1}\D R^2 + R^2 \D  \Omega_{(n-2)}^2,
\label{eq:st}
\eea
where $M={\cal M}$ is the Arnowitt-Deser-Misner (ADM) mass.


\subsection{Einstein-Lanczos-Lovelock gravity}
Einstein-Lanczos-Lovelock gravity is a natural generalization of general relativity in arbitrary dimensions as a second-order quasilinear theory of gravity~\cite{Lanczos1938, Lovelock1971}.
The second-order field equations ensure the ghost-free nature of the theory and ELL gravity reduces to general relativity with a cosmological constant in four dimensions. (See~\cite{lovelockreview, lovelockreview2} for a review of Lovelock black holes.)

The ELL action~\cite{Lanczos1938, Lovelock1971} in vacuum is a sum of dimensionally extended Euler densities given by
\begin{align}
\label{action}
I_{\rm EL}=&\frac{1}{16\pi G_{(n)}}\int \D
^nx\sqrt{-{g}}\sum_{p=0}^{[n/2]}\alpha_{(p)}{\ma L}_{(p)},\\
{\ma L}_{(p)}:=&\frac{1}{2^p}\delta^{\mu_1\cdots \mu_p\nu_1\cdots
\nu_p}_{\rho_1\cdots \rho_p\sigma_1\cdots
\sigma_p}{\cal R}_{\mu_1\nu_1}^{\phantom{\mu_1}\phantom{\nu_1}\rho_1\sigma_1}\cdots
{\cal R}_{\mu_p\nu_p}^{\phantom{\mu_p}\phantom{\nu_p}\rho_p\sigma_p},
\end{align}
where 
\begin{align}
\delta^{\mu_1\cdots \mu_p}_{\rho_1\cdots
\rho_p}:=&p!\delta^{\mu_1}_{[\rho_1}\cdots \delta^{\mu_p}_{\rho_p]}.
\end{align}
The gravitational equation following from this action is given by
\begin{align}
{\ma G}_{\mu\nu}=0, \label{beqL}
\end{align}
where
\begin{align}
{\ma G}_{\mu\nu} :=& \sum_{p=0}^{[n/2]}\alpha_{{(p)}}{G}^{(p)}_{\mu\nu},
\label{generalG}\\
{G}^{\mu(p)}_{~~\nu}:=& -\frac{1}{2^{p+1}}\delta^{\mu\eta_1\cdots
\eta_p\zeta_1\cdots \zeta_p}_{\nu\rho_1\cdots \rho_p\sigma_1\cdots
\sigma_p}{\cal R}_{\eta_1\zeta_1}^{\phantom{\eta_1}\phantom{\zeta_1}\rho_1\sigma_1}\cdots
{\cal R}_{\eta_p\zeta_p}^{\phantom{\eta_p}\phantom{\zeta_p}\rho_p\sigma_p}.
\end{align}
${G}^{(p)}_{\mu\nu}\equiv 0$  for $p\ge [(n+1)/2]$.

As a concrete example, when $n=4$, the only  nonzero contributions come from $p=0,1,2$ and the action is
 \bea
 I_{\rm EL}= \frac{1}{16\pi G_{(4)}}\int \D^4x \sqrt{-g}\biggl\{\alpha_{(0)}+ {\cal R} + \alpha_{(2)}\left({\cal R}^2 - 4{\cal R}_{\mu\nu} {\cal R}^{\mu\nu} + {\cal R}_{\mu \nu \rho \sigma}{\cal R}^{\mu \nu \rho \sigma} \right)  \biggl\},
  \label{eq:GB action}
  \eea
where we have set $\alpha_{(1)}=1$ without loss of generality.
In four dimensions, the quadratic term is topological and does not contribute to the gravitational equations, namely ${G}^{(2)}_{\mu\nu}\equiv 0$.

Consider an $n$-dimensional spherically symmetric spacetime (\ref{eq:higherDg}).
In~\cite{Kunstatter2012, Kunstatter2013, Taves2013} it was shown that the spherically symmetric ELL action takes the form
\bea
I_{(2)}=\frac{1}{l^{n-2}}\int \D^2{y}\sqrt{-{\bar g}}R^{n-2}\sum^{[n/2]}_{p=0}\alpha_{(p)} {\cal L}_{(p)},
\label{eq:SphericalLovelockAction}
\eea
where
\begin{align}
\label{eq:LovelockSimplified}
 {\cal L}_{(p)} =& \frac{(n-2)!}{(n-2p)!} \Biggl[p\Ri[{\bar g}] R^{2-2p} 
+ (n-2p)(n-2p-1)\biggl\{\left(1-Z\right)^{p} +2pZ\biggl\}R^{-2p} \nonumber\\
&  + p(n-2p)R^{1-2p}\biggl\{1-(1-Z)^{p-1}\biggl\}(D_A R)\frac{(D^A Z)}{Z} \Biggr] 
\end{align}
and we have defined
\be
Z:=(D R)^2. \label{def-Z}
\ee

The generalized Misner-Sharp mass in ELL gravity was defined~\cite{Maeda2011} as
\begin{align}
\label{qlm}
{\cal M}:=& \frac{n-2}{l^{n-2}}\sum_{p=0}^{[n/2]}{\tilde
\alpha}_{(p)}R^{n-1-2p}[1-(DR)^2]^p,
\end{align}
where 
\begin{align}
{\tilde \alpha}_{(p)}:=&\frac{(n-3)!\alpha_{(p)}}{(n-1-2p)!}.
\label{alphatil}
\end{align}
We note that ${\tilde \alpha}_{(p)} \equiv 0$~$(p\ge 2)$ for
$n=3,4,5,\cdots, 2p-1,2p$ by definition.

The generalized Misner-Sharp mass (\ref{qlm}) satisfies $D_A{\cal M}=0$ and hence ${\cal M}$ is constant in vacuum.
The resulting Birkhoff's theorem in Lovelock gravity~\cite{zegers2005} shows that, under several technical assumptions, the unique vacuum solution is given by the following Schwarzschild-Tangherlini-type solution:
\begin{align}
\D s_{(n)}^2=&-f(R)\D t^2+f(R)^{-1}\D R^2+R^2\D \Omega_{(n-2)}^2, \label{f-vacuum}
\end{align}
where the function $f(R)$ is determined by the following algebraic equation~\cite{Wheeler1986,whitt1988}:
\begin{align}
M=& \frac{n-2}{l^{n-2}}\sum_{p=0}^{[n/2]}{\tilde \alpha}_{(p)}R^{n-1-2p}(1-f)^p,
\end{align}
where $M={\cal M}$.

\section{New 2D dilaton gravity}
\label{sec2}
A natural way to generalize the spherically symmetric action~\eqref{eq:dim_reduced_action} in Einstein gravity is 2D dilaton gravity:
\begin{align}
  \label{eq:Action2}
  I_{(2)}&=\frac{1}{l^{n-2}}\int \D^2{y} \sqrt{-{\bar g}}  \Big\{ \phi(R)\Ri[{\bar g}]  + h(R)(DR)^2 +V(R) \Big\},
\end{align}
where $\phi(R)$, $h(R)$, and $V(R)$ are arbitrary functions of a scalar field $R$.
This class of theories was extensively studied in the 1990's in the hopes that they would shed light on the conundra associated with black hole thermodynamics. (See~\cite{Grumiller2002} for an excellent review.) 

The action (\ref{eq:Action2}) is, up to reparametrizations, the most general action that contains at most two derivatives of the metric and areal radius.  
The resulting field equations are therefore trivially second-order. 
However, as shown in the action~\eqref{eq:SphericalLovelockAction}, the field equations can be second order even if the action contains the higher derivative term with $D_AZ$.
This motivates us to generalize the spherically symmetric action~\eqref{eq:SphericalLovelockAction} of Einstein-Lanczos-Lovelock gravity, as follows.

\subsection{Action and field equations}
\label{subsection:Action}

In analogy with the action (\ref{eq:Action2}), we now consider the following natural extension of the spherically symmetric ELL action (\ref{eq:SphericalLovelockAction}):
\bea
\label{eq:lagrangianL}
I_{\rm XL} 
=
\frac{1}{l^{n-2}}\int \D^2{y} \sqrt{-{\bar g}}  \Big\{ \phi(R)\Ri[{\bar g}]  +\HRZ+ \chi(R,Z) (D_A R)\frac{(D^A Z)}{Z}  \Big\}.
\eea
$\HRZ$ and $\chi(R,Z)$ are as yet arbitrary functions of a scalar field $R$ and $Z$ defined by Eq.~(\ref{def-Z}).

Eq.~(\ref{eq:lagrangianL}) is the starting point for our analysis. Note that the XL action $I_{\rm XL}$ contains higher powers (potentially an infinite number of them) of $Z$ and hence of  the ``velocity'' $R_{,t}$ of the areal radius. 
We conjecture that this is the most general 2D action involving only the metric and a scalar that yields second-order equations for both\footnote{Recently Tibrewala~\cite{Tibrewala2015} used Hamiltonian techniques involving loop quantum gravity motivated variables to construct a set of new second derivative spherically symmetric gravity theories. Since the action was not written in covariant form it is difficult to say for sure whether or not it falls into the class described by (\ref{eq:lagrangianL}). }.
Moreover, we will now show that for any given $\phi(R)$ and $\chi(R,Z)$, one can choose the function $\HRZ$ so that the field equations obey Birkhoff's theorem, i.e., there is a unique one parameter family of solutions that admit at least one Killing vector.

Here we note that our approach is to think of 2D dilaton gravity as deformed spherically symmetric gravity/Lovelock gravity and identify from the beginning the geometrical quantities ${\bar g}_{AB}$ and $R$ that correspond to the metric and areal radius in the higher dimensional theory. This leads to the general form of the 2D dilaton action (\ref{eq:lagrangianL}). We could for example identify some function $F(R)$ with the higher dimensional areal radius, but this would just change the definitions of the arbitrary functions in the action.

We now present the field equations.
Varying the action (\ref{eq:lagrangianL}) for $g_{AB}$ and $R$ gives
\bea
\label{eq:covg}
0&=& (\chi-\phi_{,R})(D_A D_B R - g_{AB} D^2 R)
 + g_{AB}\biggl(\phi_{,RR}Z -\frac{1}{2} \eta\biggl) \nonumber\\
&&+ (-\phi_{,RR} + \eta_{,Z} -\chi_{,R}) (D_A R)(D_B R)
\eea
and 
\bea
\label{eq:covR}
0&=& -(\chi-\phi_{,R})\Ri +\eta_{,R} + 2(\chi_{,RR}-\eta_{,RZ})Z + (4\chi_{,R} - 2\eta_{,Z})D^2R  \nonumber \\
&&+2\chi_{,Z} [(D^2R)^2 - (D_AD_BR)(D^AD^BR)] +2(\chi_{,RZ} - \eta_{,ZZ})(D_AZ)(D^AR),
\eea
respectively, where a comma denotes the partial derivative.
These equations are clearly second order.
To obtain Eq.~(\ref{eq:covg}), it is useful to use the identity (A.6) in~\cite{Maeda2008} but with $r \to R^2$.  
The derivation of Eq.~(\ref{eq:covR}) requires the identities (A.5) and (A.8) in~\cite{Taves2013}.

\subsection{Mass function}
\label{sec:MassFunction}
We have shown that our new theory (\ref{eq:lagrangianL}) gives rise to the second-order field equations (\ref{eq:covg}) and (\ref{eq:covR}).
In order to make the theory resemble Einstein-Lanczos-Lovelock gravity as closely as possible, we require that it admit a generalized Misner-Sharp mass ${\cal M}$ as the first integral of the field equations. In ELL gravity, the mass function satisfies
\begin{align}
D_A{\cal M}={\cal G}_{AB}(D^B R)-{\cal G}^B_{~B}(D_AR),  \label{DM}
\end{align}
where the gravitational equations take the form  ${\cal G}_{AB}=0$, guaranteeing that the mass function is constant on shell.

The mass function ${\cal M}$ satisfying Eq.~(\ref{DM}) has a physical interpretation as quasi-local mass.
In ELL gravity with a minimally coupled matter field, the gravitation equations in $n$ dimensions are given by ${\cal G}_{\mu\nu}=8\pi G_{(n)} T_{\mu\nu}$, where the gravitational tensor ${\cal G}_{\mu\nu}$ is defined by Eq.~(\ref{generalG}) and $T_{\mu\nu}$ is the energy-momentum tensor for the matter field.
Now we consider the following energy-momentum;
\begin{align}
T^\mu_{~\nu}=\mbox{diag}(-\rho,P_{\rm r},P_{\rm t},\cdots,P_{\rm t}),
\end{align}
where $\rho$, $P_{\rm r}$, and $P_{\rm t}$ are energy density, radial pressure, and tangential pressure, respectively.
Then, in the following coordinates;
\begin{align}
\D s^2=g_{tt}(t,r)\D t^2+g_{rr}(t,r)\D r^2+R(t,r)^2\D \Omega_{(n-2)}^2,
\end{align}
the component $A=r$ of Eq.~(\ref{DM}) gives
\begin{align}
{\cal M}_{,r}={\cal A}_{(n-2)}\rho R^{n-2}R_{,r}.
\end{align}
The above equation gives a physical interpretation of ${\cal M}$ as quasi-local mass inside the areal radius $R$ on a spacelike hypersurface with constant $t$:
\begin{align}
{\cal M}=\int {\cal A}_{(n-2)}\mu R^{n-2}\frac{\partial R}{\partial r}\D r.
\end{align}
For this reason, we expect Eq.~(\ref{DM}) to provide a suitable definition for  the mass function ${\cal M}$.

In the present case, the two-tensor ${\cal G}_{AB}$ is obtained directly from (\ref{eq:covg}):
\bea
{\cal G}_{AB}&:=& 2(\chi-\phi_{,R})(D_A D_B R - g_{AB} D^2 R)
 + 2g_{AB}\biggl(\phi_{,RR}Z -\frac{1}{2} \eta\biggl) \nonumber\\
&&+ 2(-\phi_{,RR} + \eta_{,Z} -\chi_{,R}) D_A R D_B R
\eea
so that
\begin{align}
{\cal G}_{AB}(D^B R)-{\cal G}^B_{~B}(D_AR) 
 =& (\chi-\phi_{,R})D_AZ- 2\biggl(\phi_{,RR}Z -\frac{1}{2} \eta\biggl)D_A R. \label{int1}
\end{align}
In order for the theory to have a mass function that obeys (\ref{DM}), 
Equation~(\ref{int1}) shows that ${\cal M}={\cal M}(R,Z)$ must satisfy the following:
\begin{align}
\frac{\partial {\cal M}}{\partial Z}=&\chi-\phi_{,R}, \label{dMdZ}\\
\frac{\partial {\cal M}}{\partial R}=&- 2\phi_{,RR}Z + \eta. \label{dMdR}
\end{align}
The necessary and sufficient condition for the existence of such a mass function is therefore the  integrability condition $\partial^2 {\cal M}/\partial R\partial Z=\partial^2 {\cal M}/\partial Z\partial R$, which gives the following constraint on the Lagrangian functions:
\begin{align}
\phi_{,RR}=\eta_{,Z}-\chi_{,R}. \label{eq:integrability}
\end{align}
 Straightforward integration of (\ref{int1}) yields the following two integral forms for the mass function:
\begin{align}
{\cal M}:=-\phi_{,R}Z+\int^Z\chi(R,{\bar Z})\D{\bar Z}+{\cal M}_0(R) \label{qlm-covariant} 
\end{align}
and 
\begin{align}
{\cal M}:=- 2\phi_{,R}Z +\int^R\eta({\bar R},Z)\D{\bar R} +{\cal M}_1(Z). \label{qlm-covariant2}
\end{align}
In order to uniquely determine the correct expression for the mass function ${\cal M}{(R,Z)}$, one must in general evaluate both integrals and fix the two arbitrary functions appropriately.
The integrability conditions guarantee that there will always be a choice of ${\cal M}_0(R)$ and ${\cal M}_1(Z)$ to make the expressions for the two expressions
(\ref{qlm-covariant}) and (\ref{qlm-covariant2}) consistent. 
For example, in Einstein-Gauss-Bonnet gravity, which is second-order ELL gravity, we have
\begin{align}
\phi(R)=&R^{n-2}+2(n-2)(n-3){\alpha}_{(2)}R^{n-4},\\
\chi(R,Z)=&2(n-2)(n-3)(n-4){\alpha}_{(2)}R^{n-5}Z,\\
\eta(R,Z)=&(n-2)(n-3)R^{n-4}(1+Z) \nonumber \\
&+(n-2)(n-3)(n-4)(n-5){\alpha}_{(2)}R^{n-6}(1+Z)^2.
\end{align}
Integration of Eq.~(\ref{qlm-covariant2}) gives the correct mass function~\cite{Maeda2008} with ${\cal M}_1 =0$:
\begin{align}
  {\cal M}=(n-2)R^{n-3}(1-Z)+(n-2)(n-3)(n-4){\alpha}_{(2)}R^{n-5}(1-Z)^2.
  \label{eq:IntM1}
\end{align}
In contrast, since $\eta(R,Z)$ contains terms depending only on $R$, Eq.~(\ref{qlm-covariant}) gives the following:
\begin{align}
{\cal M}=&(n-2)R^{n-3}(1-Z)+(n-2)(n-3)(n-4){\alpha}_{(2)}R^{n-5}(1-Z)^2 \nonumber \\
&-(n-2)R^{n-3}-(n-2)(n-3)(n-4){\alpha}_{(2)}R^{n-5}+ {\cal M}_0(R).
\label{eq:IntM2}
\end{align}
Consistency between (\ref{eq:IntM1}) and (\ref{eq:IntM2}) requires the choice:
\be
{\cal M}_0(R)=(n-2)R^{n-3}+(n-2)(n-3)(n-4){\alpha}_{(2)}R^{n-5}.
\ee

\subsection{Birkhoff's theorem}
We now show that the theory (\ref{eq:lagrangianL}) with the condition (\ref{eq:integrability}) obeys Birkhoff's theorem.
Here we assume $Z=(DR)^2\ne 0$ for simplicity.
Then we can  choose without loss of generality $R$ as a coordinate such that 
\begin{align}
\D s^2 =& -f(t,R)e^{2\delta(t,R)}\D t^2+f(t,R)^{-1}\D R^2. \label{metric-birkhoff}
\end{align}
In this case, the components of the gravitational tensor ${\cal G}^A_{~B}$ are 
\begin{align}
{\cal G}^t_{~t}-{\cal G}^R_{~R}=& f\delta_{,R} (\chi-\phi_{,R})- f(-\phi_{,RR} + \eta_{,Z} -\chi_{,R}),\label{vacuumeq1} \\
2{\cal G}^t_{~t}=& 2\phi_{,RR}f -\eta-f_{,R} (\chi-\phi_{,R}),\label{vacuumeq2} \\
{\cal G}^t_{~R}=& -\frac12f^{-2}f_{,t}e^{-2\delta}(\chi-\phi_{,R}), \label{vacuumeq3} \\
{\cal G}^R_{~t}=& \frac12 f_{,t}(\chi-\phi_{,R}),\label{vacuumeq4} 
\end{align}
where $\eta$ and $\chi$ and their derivatives have been evaluated at $Z=f(t,R)$.
Eq.~(\ref{eq:covR}) is an auxiliary equation in general.
We will also assume that $\chi-\phi_{,R} \ne 0$ so that vanishing of the components (\ref{vacuumeq3}) or (\ref{vacuumeq4}) imply that $f=f(R)$.

We also see that the integrability condition (\ref{eq:integrability}) makes the last term in Eq.~(\ref{vacuumeq1}) vanish, which implies $\delta=\delta(t)$.
Since we can set $\delta(t)=0$ without loss of generality by redefinition of $t$, the metric reduces to the usual Schwarzschild form:
\begin{align}
\label{eq:BasicMetric}
\D s^2 =& -f(R)\D t^2+f(R)^{-1}\D R^2.
\end{align}
The metric function $f(R)$ can now be determined algebraically from either Eq.~(\ref{qlm-covariant}) or (\ref{qlm-covariant2}), namely
\begin{align}
{\cal M}=-\phi_{,R}f+\int^f\chi(R,{\bar f})\D{\bar f} +{\cal M}_0(R)
\end{align}
or
\begin{align}
{\cal M}=- 2\phi_{,R}f +\int^R\eta({\bar R},f)\D{\bar R} +{\cal M}_1(f), 
\end{align}
where we used $Z=f$.

\subsection{Birkhoff's theorem with integrating factor}
\label{sec:factor}

In the previous two subsections, we have seen that given the integrability condition (\ref{eq:integrability}), we can define the mass function (\ref{qlm-covariant}) or (\ref{qlm-covariant2}) which in turn yields the unique, static vacuum solution given by Eq.~(\ref{eq:BasicMetric}).
We now show that even if the Lagrangian functions do not satisfy the integrability condition (\ref{eq:integrability}), we can define a quantity analogous to the mass function that is constant on shell and leads to a unique and static vacuum solution. 
The form of the solution is, however, different from the standard Schwarzschild form, in that $g_{00}g_{11}\neq -1$.
Moreover, this first integral of the field equations does not have the physical interpretation as the quasi-local mass, discussed after Eq.~(\ref{DM}) above.

We define the quantity ${\bar{\cal M}}$ by introducing an integrating factor $\Omega(R,Z)$ such that 
\begin{align}
D_A{\bar{\cal M}}=&\Omega(R,Z)\{{\cal G}_{AB}(D^B R)-{\cal G}^B_{~B}(D_AR)\} \nonumber \\
=&\Omega(R,Z)\biggl\{ (\chi-\phi_{,R})D_AZ- 2\biggl(\phi_{,RR}Z -\frac{1}{2} \eta\biggl)D_A R\biggl\}.  \label{DM-2}
\end{align}
Assuming that the integrating factor does not vanish, ${\bar{\cal M}}$ is  constant on shell.

In this case, Eq.~(\ref{DM-2}) requires ${\bar{\cal M}}={\bar{\cal M}}(R,Z)$ to satisfy
\begin{align}
\frac{\partial {\bar{\cal M}}}{\partial Z}=&\Omega(\chi-\phi_{,R}), \label{dMdZ-2}\\
\frac{\partial {\bar{\cal M}}}{\partial R}=&\Omega(- 2\phi_{,RR}Z + \eta) \label{dMdR-2}
\end{align}
and the integrability condition is now
\begin{align}
\phi_{,RR}+\chi_{,R}- \eta_{,Z}=\frac{\Omega_{,Z}}{\Omega}(- 2\phi_{,RR}Z + \eta)-\frac{\Omega_{,R}}{\Omega}(\chi-\phi_{,R}).  \label{eq:integrability2}
\end{align}
This is is a first-order quasilinear partial differential equation for $\ln|\Omega|$, so that the integrating factor $\Omega(R,Z)$ always exists for given $\phi(R)$, $\chi(R,Z)$, and $\eta(R,Z)$ with sufficient differentiability. It is not, however, easy to find analytically in general.

Finally, given the integrability condition (\ref{eq:integrability2}), ${\bar{\cal M}}$ can be written in either of the two following integral forms:
\begin{align}
{\bar{\cal M}}:=\int^Z\Omega(\chi-\phi_{,R})\D{\bar Z}+{\bar{\cal M}}_0(R) \label{qlm-covariant-2} 
\end{align}
or
\begin{align}
{\bar{\cal M}}:=\int^R\Omega({\bar R},Z)(- 2\phi_{,{\bar R}{\bar R}}Z + \eta({\bar R},Z))\D{\bar R} +{\bar{\cal M}}_1(Z). \label{qlm-covariant2-2}
\end{align}
Similar computations to those for the mass function without the integration factor (\ref{qlm-covariant}) or (\ref{qlm-covariant2}) show that one can always find integration ``constants''
$\bar{{\cal M}}_0(R)$ and $\bar{\cal M}_1(Z)$ that insure consistency of (\ref{qlm-covariant-2}) and (\ref{qlm-covariant2-2}).


We now derive the vacuum solution.
Starting from the metric (\ref{metric-birkhoff}), either Eq.~(\ref{vacuumeq3}) or (\ref{vacuumeq4}) again implies that $f=f(R)$.
The metric function $f(R)$ is again determined algebraically by Eq.~(\ref{qlm-covariant-2}) or (\ref{qlm-covariant2-2}), namely
\begin{align}
{\bar{\cal M}}=\int^f\Omega(R,{\bar f})(\chi(R,{\bar f})-\phi_{,R})\D{\bar f}+{\bar{\cal M}}_0(R)  
\end{align}
or
\begin{align}
{\bar{\cal M}}=\int^R\Omega({\bar R},f)(- 2\phi_{,{\bar R}{\bar R}}f + \eta({\bar R},f))\D{\bar R} +{\bar{\cal M}}_1(Z)  .
\end{align}
The remaining metric function $\delta(t,R)$ is determined by Eq.~(\ref{vacuumeq1}):
\begin{align}
\delta_{,R} =\frac{\phi_{,RR} - \eta_{,Z} +\chi_{,R}}{\phi_{,R}-\chi}\biggl|_{Z=f(R)}. \label{eq-delta}
\end{align}
Since the right-hand side depends only on $R$, the solution is
\begin{align}
\delta(t,R) =\int^R\frac{\phi_{,{\bar R}{\bar R}} - \eta_{,Z} +\chi_{,{\bar R}}}{\phi_{,{\bar R}}-\chi}\biggl|_{Z=f({\bar R})}\D {\bar R}+\delta_2(t). \label{sol-delta}
\end{align}
Since $\delta_2(t)$ can be set to zero by redefinition of $t$, the unique vacuum solution is again static.

In summary, under the sole assumption $\chi-\phi_{,R} \ne 0$, for a given solution $\Omega(R,Z)$ of the partial differential equation (\ref{eq:integrability2}), the metric function $f(R)$ and $\delta(R)$ are given by Eq.~(\ref{qlm-covariant-2}) (or Eq.~(\ref{qlm-covariant2-2})) and Eq.~(\ref{sol-delta}) with $\delta_2(t)=0$, respectively.

\subsection{Maximally symmetric vacua}

We now derive the general conditions on the functions $\phi(R)$, $\chi(R,Z)$, and $\eta(R,Z)$ in the action in order that the theory  admit  maximally symmetric vacuum solutions. We do this by simply inserting the vacuum solution into the field equations (\ref{vacuumeq1} - \ref{vacuumeq4}).
Adopting coordinates such that 
\bea
\D s^2&=& -(1-\lambda R^2)\D t^2+(1-\lambda R^2)^{-1}\D R^2, 
\eea
where $\lambda$ is the effective cosmological constant, the field equations give
\bea
0&=&- 2\lambda R(\chi-\phi_{,R})+ \eta +2 (1-\lambda R^2) (\chi_{,R}-\eta_{,Z}),\label{vacua1}\\
0&=& -2\lambda (\chi-\phi_{,R}) +\eta_{,R} + 2(1-\lambda R^2)(\chi_{,RR}-\eta_{,RZ}) -2\lambda R (4\chi_{,R}- 2\eta_{,Z})  \nonumber \\
&&+4\lambda^2R^2\chi_{,Z}  -4\lambda R(1-\lambda R^2)(\chi_{,RZ} - \eta_{,ZZ}),\label{vacua2}
\eea
where $\chi$, $\eta$ and their derivatives are evaluated at $Z=1-\lambda R^2$.

For the Minkowski vacuum ($\lambda=0$), the equations above reduce to the following single condition:
\bea
0= \eta +2  (\chi_{,R}-\eta_{,Z}),
\label{Mink1}
\eea
where $\chi$, $\eta$ and their derivatives are evaluated at $Z=1$. If the integrability condition is satisfied, Eq.~(\ref{Mink1}) implies that:
\be
\eta(R,1) = 2\phi_{,RR}.
\label{necessary-Minkowski2}
\ee

We now consider the (A)dS vacuum ($\lambda\ne 0$).
In this case, differentiating Eq.~(\ref{vacua1}) with respect to $R$ and using Eq.~(\ref{vacua2}), we obtain a necessary condition for existence of the (A)dS vacuum:
\begin{align}
\phi_{,RR}=\eta_{,Z}-\chi_{,R}. \label{necessary-AdS}
\end{align}
This is the same as the integrability condition (\ref{eq:integrability}) for the mass function but evaluated at $Z=1-\lambda R^2$.
If this condition is satisfied, the (A)dS vacua can be obtained for lagrangian functions $\chi$ and $\phi$ that satisfy  equation~(\ref{vacua1}), namely
\bea
0=- 2\lambda R(\chi-\phi_{,R})+ \eta -2 (1-\lambda R^2) \phi_{,RR},
\label{necessary-AdS2}
\eea
where we used Eq.~(\ref{necessary-AdS}) and $\chi$, $\eta$ are evaluated at $Z=1-\lambda R^2$. Notice that when $\lambda=0$ (\ref{necessary-AdS2}) yields  (\ref{necessary-Minkowski2}).


\section{Hamiltonian formalism}
\label{sec:hamilton}
In this section, we perform the Hamiltonian analysis of our new 2D dilaton gravity (\ref{eq:lagrangianL}), leaving the computational details to the appendix.
The Hamiltonian analysis of spherically symmetric Einstein-Gauss-Bonnet gravity was done in~\cite{Louko1997} and again using a different formalism in~\cite{Taves2012}. The Hamiltonian analysis of spherically symmetric ELL gravity was discussed in~\cite{Deser2005, Teitelboim1987a}. The following is based in large part on the methodology and results of~\cite{Kunstatter2012, Kunstatter2013}
and uses  the notation and conventions of~\cite{Taves2014}.

\subsection{ADM decomposition}

For convenience we separate the XL action (\ref{eq:lagrangianL}) into two parts
\bea
\label{eq:SplitL}
I_{\rm XL} 
 &=& I_{\rm G} +  I_{\rm L},
\eea
where 
\begin{align}
  \label{eq:Action3G}
  I_{\rm G}&:=\frac{1}{l^{n-2}}\int \D^2{y} \sqrt{-{\bar g}}  \Big\{ \phi(R)\Ri[{\bar g}]  +\HRZ \Big\},\\
    I_{\rm L} &:= \frac{1}{l^{n-2}}\int\D^2{y} \sqrt{-{\bar g}} \biggl\{ \chi(R,Z)(D_A R)\frac{(D^A Z)}{Z} \biggl\}.
\end{align}
$\HRZ$ and $\chi(R,Z)$ are as yet arbitrary functions of a scalar field $R$ and $Z$ defined by Eq.~(\ref{def-Z}).

We henceforth assume that $\chi(R,Z)$ has an expansion of the form
\be
\chi(R,Z):=\sum_I\Ialpha(R) \IW(Z) Z,
\label{eq:chi}
\ee
where $\Ialpha(R)$ and $\IW(Z)$ are arbitrary functions of $R$ and $Z$, respectively.
This assumption gives
\begin{align}
 I_{\rm L} &= \sum_I  I^{(I)}_{\rm L},
\end{align}
where
\begin{align}
I^{(I)}_{\rm L}  &:=\frac{1}{l^{n-2}}\int \D^2{y} \sqrt{-{\bar g}} \Big\{\Ialpha(R) \IW(Z) (D_A R)(D^A Z)\Big\}. \label{def-ILI}
\end{align}

To derive the Hamiltonian, we start with the general ADM metric in two spacetime dimensions:
\begin{equation}
  \label{eq:generalds2G}
  \D s^2 = -N(t,x)^2\D t^2 + \Lambda(t,x)^2(N_r(t,x)\D t + \D x)^2.
\end{equation}
The unit normal to the spacelike hypersurface with constant $t$ is given by 
\begin{equation}
u^A\frac{\partial}{\partial {y}^A}=N^{-1}\biggl(\frac{\partial}{\partial t}-N_r\frac{\partial}{\partial x}\biggl)=:\frac{\partial}{\partial u}.\label{defy}
\end{equation}
  
In this parametrization we have\footnote{The following is a slight departure from the notation in~\cite{Taves2014}, where $y$ was used rather than $u$ for $R_{,u}$.}
\begin{align}
\label{eq:ZG}
&Z =-{R_{,u}}^2+\Lambda^{-2}{R_{,x}}^2,
\end{align} 
where the operator $\partial/\partial u$ is defined by Eq.~(\ref{defy}).
Using this notation, the action (\ref{eq:Action3G}) becomes
\begin{align}
  \label{eq:Action4G}
  I_{\rm G}
& = \frac{2}{l^{n-2}}\int \D t\D x 
      N\left\{\phi_{,R}R_{,u}\left(-\Lambda_{,u} +\frac{N_{r,x}}{N}\Lambda\right) -\left(\frac{\phi_{,x}}{\Lambda}\right)_{,x} +\frac{1}{2} \HRZ \Lambda\right\}.
\end{align}
The contributions from $I_G$ to the momenta conjugate to $\Lambda$ and $R$ are given, respectively, by
\begin{align}
  \label{eq:PLambda1}
  P^{\rm (G)}_\Lambda =&  -2l^{-(n-2)}\phi_{,R} R_{,u},\\
  \label{eq:PR1}
  P^{\rm (G)}_R =& 2l^{-(n-2)}\left(-\Lambda_{,u} \phi_{,R} + \frac{N_{r,x}}{N}\Lambda \phi_{,R} -{R_{,u}} \Lambda\frac{\partial \HRZ }{\partial Z}\right).
\end{align}
We note for future reference that
\be
\delta{Z} = -2R_{,u}\delta R_{,u} + \delta b,
\ee
where we have defined for convenience
\be
b:= \frac{{R_{,x}}^2}{\Lambda^2}.
\ee

In order to write $I^{(I)}_{\rm L}$ defined by Eq.~(\ref{def-ILI}) in terms of phase space variables, 
we assume that $\IW(Z)(=\IW(-{R_{,u}}^2+b))$ has a Taylor expansion in $Z$ and hence in ${R_{,u}}^2$, so that
\be
\IW(Z)=\sum_{m} \Hb {R_{,u}}^{2m},
\ee
where
\be
\Hb := \frac{(-1)^m}{m ! }\left. \frac{\D^m W_{(I)}}{\D Z^m}\right|_{Z=b}.
\ee
This gives us the relation:
\bea
\IW(Z)R_{,u}\delta Z &=& \IW(Z)R_{,u}(-2R_{,u}\delta R_{,u}+\delta b)\nonumber\\
    &=&-2 \sum_{m=0} \Hb {R_{,u}}^{2m+2}\delta R_{,u}  + \IW(Z) R_{,u}\delta b\nonumber\\
   &=& -2\sum_{m=0} \Hb \frac{\delta ({R_{,u}}^{2m+3})}{2m+3}  +\IW(Z) R_{,u}\delta b.
\eea

Using the above, after another lengthy derivation (see Appendix~\ref{app:Lagrangian}), we obtain the Lagrangian density ${\cal L}^{(I)}_{\rm L}$ for each of the terms satisfying $I^{(I)}_{\rm L}=\int \D^2{y} \sqrt{-{\bar g}} {\cal L}^{(I)}_{\rm L}$: 
\bea
l^{n-2} {\cal L}^{(I)}_{\rm L}&=&l^{n-2}\left({} P^{(I)}_\Lambda \Lambda_{,t} + \Lambda N_r{P^{(I)}_{\Lambda}}_{,x} 
    -(N_r{} P^{(I)}_\Lambda \Lambda)_{,x}\right)\nonumber\\
& &  +2NR_{,u}\left( \Ialpha(R)\IW(Z)R_{,u} \frac{R_{,x}}{\Lambda} + 
      2\Ialpha(R) \frac{R_{,x}}{\Lambda} \sum_m \frac{\D \Hb }{\D b}\frac{{R_{,u}}^{2m+3}}{2m+3}
      \right)_{,x}\nonumber\\
 & &  -2N\Ialpha_{,R} \Lambda R_{,u} \sum_m \Hb \frac{{R_{,u}}^{2m+3}}{2m+3}
  +\frac{N}{\Lambda}\Ialpha(R){R_{,x}}{\IW(Z)}Z_{,x},
\label{eq:IDeltaIFinal}
\eea
where we have dropped total divergences and defined
   \begin{align}
{} P^{(I)}_\Lambda:=&
\frac{1}{l^{n-2}}\biggl\{- 2\sum_{m}\Ialpha(R) \left( \Hb -2 \frac{\D \Hb }{\D b}\frac{{R_{,x}}^2}{\Lambda^2}\right)\frac{{R_{,u}}^{2m+3}}{2m+3}
  +2\Ialpha(R) \IW(Z)R_{,u}\frac{{R_{,x}}^2}{\Lambda^2}   \biggl\}.
\label{eq:IDeltaPLambda}
\end{align}

A key feature of this class of theories is that the total action is linear in ${\Lambda}_{,t}$. 
Thus from the above it is easy to extract the expression for the total $P_\Lambda$:
\be
P_\Lambda = -\frac{2}{l^{n-2}} \phi_{,R}R_{,u} +\sum_I {} P^{(I)}_\Lambda.
\label{eq:FullPLambda}
\ee
Equation~(\ref{eq:FullPLambda}) provides implicitly an expression for $R_{,u}$ as a function of the other phase space variables. From (\ref{eq:FullPLambda}) we see that $P_R$, the momentum conjugate of $R$, is conspicuously absent in the final expression for $R_{_,u}$, which therefore depends only on $\Lambda$, $P_\Lambda$, and $R$. The expression for $P_R$ analogous to (\ref{eq:FullPLambda}) is significantly more complicated but another important aspect of spherically symmetric ELL gravity that is shared by our XL gravity is that one does not need to make use of this expression when deriving the Hamiltonian equations of motion.

The total Hamiltonian density is
\bea
{\cal H}_{\rm XL} &=&  P_{\Lambda} \Lambda_{,t} + P_R R_{,t} - {\cal L}_{\rm XL}\nonumber\\
   &=& N {\cal H} + N_r {\cal H}_r,
\label{eq:HT}
\eea
where the total Lagrangian density is 
\bea
\label{eq:totalL}
{\cal L}_{\rm XL}:=
\frac{1}{l^{n-2}}\Big\{ \phi(R)\Ri  + \HRZ + \chi(R,Z) (D_A R)\frac{(D^A Z)}{Z}  \Big\}.
\eea
The terms linear in $\Lambda_{,t}$ again cancel, leaving:
\bea
{\cal H}&:=& P_R R_{,u}   +\frac{1}{l^{n-2}}\Biggl[2\left(\frac{\phi_{,x}}{\Lambda}\right)_{,x}-{\Lambda} \HRZ   \nonumber\\
  & & -2R_{,u}\sum_I\left( \Ialpha(R)\IW(Z)R_{,u} \frac{R_{,x}}{\Lambda} + 
      2\Ialpha(R) \frac{R_{,x}}{\Lambda} \sum_m \frac{\D \Hb}{\D b}\frac{{R_{,u}}^{2m+3}}{2m+3}
      \right)_{,x}\nonumber\\
 & &  +2\Lambda R_{,u}\sum_I\Ialpha_{,R}  \sum_m \Hb \frac{{R_{,u}}^{2m+3}}{2m+3}
    - \sum_I\frac{R_{,x}}{\Lambda}\Ialpha(R){\IW(Z)}Z_{,x} \Bigg], \\
{\cal H}_r &:=& P_R R_{,x} - {P_\Lambda}_{,x}\Lambda.
\eea
Finally, the ADM form of the XL action (\ref{eq:lagrangianL}) is given by 
\bea
I_{\rm XL} 
&=&
\int \D t \D x\biggl(\Lambda_{,t}P_\Lambda+R_{,t}P_R-N{\cal H}-N_r{\cal H}_r\biggl).
\eea
Variation of the Lagrange multipliers $N$ and $N_r$ give the Hamiltonian constraint ${\cal H}=0$ and the diffeomorphism constraint ${\cal H}_r=0$, respectively.
Note that since $R_{,u}$ is an implicit function of $P_\Lambda$, $\Lambda$, and $R$, the momentum conjugate to $R$, namely $P_R$, appears only in the first terms of both the Hamiltonian and diffeomorphism constraints.

\subsection{Mass function}

The general Hamiltonian procedure for obtaining the mass function is to take the linear combination of ${\cal H}$ and ${\cal H}_r$ that eliminates $P_R$:
\begin{align}
\tilde{\cal H} :=& \frac{R_{,x}}{ \Lambda}{\cal H} -\frac{R_{,u}}{ \Lambda}{\cal H}_r \nonumber\\
    =& R_{,u}{P_\Lambda}_{,x} +\frac{1}{l^{n-2}}\Bigg[\frac{2R_{,x}}{\Lambda}\left(\frac{\phi_{,x}}{\Lambda}\right)_{,x}  - \HRZ R_{,x}\nonumber\\
   & -2R_{,u}\frac{R_{,x}}{\Lambda}\sum_I\left\{ \Ialpha(R)\IW(Z)R_{,u} \frac{R_{,x}}{\Lambda} + 
      2\Ialpha(R) \frac{R_{,x}}{\Lambda} \sum_m \frac{\D \Hb }{\D b}\frac{{R_{,u}}^{2m+3}}{2m+3}
      \right\}_{,x}\nonumber\\
  &  +2 R_{,x} R_{,u} \sum_I \left\{\Ialpha_{,R} \sum_m \Hb \frac{{R_{,u}}^{2m+3}}{2m+3}  
  - \left(\frac{R_{,x}}{\Lambda}\right)^2\Ialpha(R){\IW(Z)}Z_{,x}
\right\} \Bigg].
\label{eq:ITildeH1}
\end{align}
By expanding the explicit expression for $P_\Lambda$ (see Appendix~\ref{app:Hamiltonian}), we obtain
\bea
\tilde{\cal H} =\biggl(2\phi_{,RR}Z- \HRZ \biggl) R_{,x} + \biggl(\phi_{,R}-\chi(R,Z)\biggl) Z_{,x}.
\label{eq:IFinalH}
\eea
The total Hamiltonian is now
\be
H  = \int \D x \left(\tilde{N}\tilde{\cal H} + \tilde{N}_r {\cal H}_r\right) +H_{\rm B},
\label{eq:H1}
\ee
where $H_{\rm B}$ is the boundary term required to make the variational principle well defined and will be made explicit below.
New Lagrange multipliers $\tilde{N}$ and $\tilde{N}_r$ are defined by 
\bea
\tilde{N} &:=&   \frac{N \Lambda}{R_{,x}},   \\
\tilde{N}_r &:=&  N_r + N\frac{R_{,u}}{R_{,x}},   
\eea
whose variations give the constraint equations $\tilde{\cal H}=0$ and ${\cal H}_r=0$, respectively.

Given the form of the Hamiltonian density in (\ref{eq:IFinalH}), the procedure for finding a mass function ${\cal M}(R,Z)$ follows as in Section~\ref{sec:MassFunction}. In the Hamiltonian context, we have
\bea
\tilde{\cal H} &=& \biggl(2\phi_{,RR}Z- \HRZ \biggl) R_{,x} + \biggl(\phi_{,R}-\chi(R,Z)\biggl) Z_{,x}\nonumber\\
  & =& - {\cal M}_{,x},
\label{eq:dM}
\eea
so that ${\cal M}= {\cal M}(t)$ is satisfied on the constraint surface.
From the expression ${\cal M}_{,x}={\cal M}_{,R}R_{,x}+ {\cal M}_{,Z} Z_{,x} $, we obtain the same integrability condition for the existence of such a mass function as before, namely (\ref{eq:integrability}). 
Finally, we obtain the mass function defined by Eq.~(\ref{qlm-covariant}) or (\ref{qlm-covariant2}).
It is also straightforward to verify that the mass function commutes with the total Hamiltonian, and therefore is also independent of time.

Lastly, we derive the boundary term $H_{\rm B}$ in the asymptotically flat case. The variation of the total Hamiltonian $H$ takes the form
\bea
\delta H &=& \int \D x \left(-\tilde{N}\delta {\cal M}_{,x} + \tilde{N}_r \delta {\cal H}_r\right) +\delta H_{\rm B} \nonumber\\
&=& \int \D x \left\{\tilde{N}_{,x}\delta {\cal M} + \tilde{N}_r \delta {\cal H}_r -\left(\tilde{N}\delta {\cal M}\right)_{,x}\right\} + \delta H_B.
\eea
We have neglected the variation of the Lagrange multipliers, which merely enforces the constraints. In the asymptotically flat case, we assume $\tilde{N}\to 1$ at infinity, while ${\cal M}=M=$constant for vacuum solutions. Thus, as anticipated, the required boundary term is
\be
H_{\rm B} = \int \D x\left(\tilde{N}{\cal M}\right)_{,x}\biggl|_{x=x_{\rm B}} = M,
\label{eq:HB}
\ee
which is the ADM mass, where $x=x_{\rm B}$ corresponds to the asymptotically flat region.
Moreover, $\tilde{N}_r\to 0$ holds in the asymptotically flat case so that this is in fact the only boundary term required.

\section{Designing nonsingular black holes}
\label{sec5}
In this section, we show how to construct specific nonsingular black holes as exact solution by making appropriate choices for the functions in the XL action (\ref{eq:lagrangianL}).
We are interested in constructing ``physical" nonsingular black holes whose curvature is everywhere bounded for arbirarily large $M$.

\subsection{Criterion for physical nonsingular black holes}
The standard 2D dilaton gravity theory (\ref{eq:Action2}) is solvable and obeys Birkhoff's theorem. By making suitable choices of the functions of $R$ in the action, the solutions can describe nonsingular spacetimes that have either two horizons or, at least in one special case, one horizon. 
If one chooses, for example, $h(R) = V(R) = \phi_{,RR}(R)$, the most general solution is~\cite{Taves2014}
\begin{align}
\D s^2:=&{\bar g}_{AB}\D {y}^A \D {y}^B \nonumber \\
=&-\left(1-\frac{l^{n-2}{M}}{j(R)}\right)\D t^2 +  \left(1-\frac{l^{n-2}{M}}{j(R)}\right)^{-1}\D R^2, \label{eq:ds2vacuum}
\end{align}
where 
\begin{align}
j(R)&:=\int  V(R) \D R.
\label{eq:2Dsoln}
\end{align}
As expected this solution contains a single parameter, $M$, and has at least one Killing vector $\partial/\partial t$. There are Killing horizons whenever $j=l^{n-2}M$, and the Killing vector is timelike in the asymptotic region, $j>l^{n-2}M$. 

A special case of particular interest when $n=4$ is $\phi_{,R} = j(R) = (R^2+l^2)^{3/2}/R^2$, which produces the well-known Bardeen metric~\cite{Bardeen1968}:
\be
\D s_{(4)}^2 = -\left(1- \frac{l^{2}M R^2}{(R^2+l^2)^{3/2}}\right)\D t^2 + \left(1- \frac{l^{2}M R^2}{(R^2+l^2)^{3/2}}\right)^{-1}\D R^2+R^2\D \Omega^2_{(2)}.
\label{eq:metric1}
\ee
Near $R=0$ this metric approaches de~Sitter spacetime with curvature of order $M/l$. It asymptotes to the Schwarzschild solution at spatial infinity as required. 
Although the Bardeen spacetime (\ref{eq:metric1}) is everywhere nonsingular, the maximum value of the curvature clearly grows without bound as the mass of the black hole is increased.
For this reason, we consider this class of nonsingular black holes to be unphysical.

Loosely speaking, the problem with the Bardeen spacetime (\ref{eq:metric1}) stems from the fact that the mass $M$ appears only in the numerator of the second term in $g_{00}$.
It turns out to be more difficult to find theories in which the mass $M$ appears in the denominator of the metric functions. In fact, to the best of our knowledge it cannot be done within the framework of the action for pure 2D dilaton gravity (\ref{eq:Action2}).

\subsection{Exact solutions}
\subsubsection{Hayward black hole}
The following Hayward nonsingular black hole~\cite{Hayward2006,Frolov2014} is different from the Bardeen black hole (\ref{eq:metric1}) in that it is physical in the sense defined above:
\begin{align}
f(R)=1-\frac{l^{2}MR^2}{R^{3}+l^4M}.
\end{align}
this metric also approaches the de Sitter form near $R=0$, but this time the curvature goes as $1/l^2$. The curvature is therefore  bounded for arbitrarily large $M$.

The Hayward black hole can be easily generalized in  $n$ dimensions as
\begin{align}
f(R)=1-\frac{l^{n-2}MR^2}{R^{n-1}+l^nM}. \label{Hayward-BH-n}
\end{align}
The mass-horizon relation is given by   
\begin{align}
M=\frac{R_{\rm h}^{n-1}}{l^{n-2}(R_{\rm h}^2-l^2)},
\end{align}
where $R=R_{\rm h}$ is the radius of the Killing horizon, defined by $f(R_{\rm h})=0$.
This shows that, for $n\ge 4$, a black-hole configuration with outer and inner horizons is realized for $M> M_{\rm ex}$, where
\begin{align}
M_{\rm ex}=\frac{n-3}{2l}\biggl(\frac{n-1}{n-3}\biggl)^{(n-1)/2}.
\end{align}
The lower bound $M=M_{\rm ex}$ gives an extremal black hole.
Unlike the Reissner-Nordstr\"om black hole, the radius of the inner horizon converges to $l$ in the limit of $M\to \infty$.

Let us identify the conditions on the function in the XL action (\ref{eq:lagrangianL}) to admit this nonsingular black hole.
 Replacing $f$ by $Z$ in Eq.~(\ref{Hayward-BH-n}), we obtain
\begin{align}
{\cal M}=\frac{R^{n-1}(1-Z)}{l^{n-2}R^2-l^n(1-Z)}
\end{align}
and hence
\begin{align}
\frac{\partial{\cal M}}{\partial R}=&\frac{(n-3)l^{n-2}R^{n}(1-Z)-(n-1)l^nR^{n-2}(1-Z)^2}{\{l^{n-2}R^2-l^n(1-Z)\}^2},\\
\frac{\partial{\cal M}}{\partial Z}=&-\frac{l^{n-2}R^{n+1}}{\{l^{n-2}R^2-l^n(1-Z)\}^2}.
\end{align}
Comparing this to Eqs.~(\ref{dMdZ}) and (\ref{dMdR}), we find the conditions:
\bea
\eta(R,Z)&=&2\phi_{,RR}Z+\frac{(n-3)l^{n-2}R^{n}(1-Z)-(n-1)l^nR^{n-2}(1-Z)^2}{\{l^{n-2}R^2-l^n(1-Z)\}^2}, \nonumber\\
\chi(R,Z)&=&\phi_{,R}-\frac{l^{n-2}R^{n+1}}{\{l^{n-2}R^2-l^n(1-Z)\}^2}.
\eea

 Note that since $Z$ takes values in the range $(0,1)$, the denominators in $\chi$ and $\eta$ above are nowhere vanishing. As well one can verify that $\phi_{,R}-\chi$ vanishes only at the point $R=0$, just as in the general relativistic limit $l=0$, so that the gauge condition is valid  everywhere except at the trivial coordinate singularity $R=0$.

\subsubsection{Bardeen-type black hole}

It is possible to construct physical nonsingular black holes similar to the Bardeen black hole (\ref{eq:metric1}) using the action $I_{\rm XL}$.
The $n$-dimensional version of this Bardeen-type black hole is  
\begin{align}
f(R)=1-\frac{l^{n-2}MR^2}{(R^2+M^{2/(n-1)}l^{2n/(n-1)})^{(n-1)/2}}, \label{Bardeen-type}
\end{align}
where we have assumed $M \ge 0$.
This metric also reduces to de~Sitter with bounded curvature for large $M$.
The mass-horizon relation for this Bardeen-type black hole is  
\begin{align}
M=\frac{R_{\rm h}^{n-1}}{l^{n-2}(R_{\rm h}^{4/(n-1)}-l^{4/(n-1)})^{(n-1)/2}}
\end{align}
and the parameter dependence for a black-hole configuration is similar to the Hayward black hole.
For this Bardeen-type black hole, the mass parameter for extremal an black hole is 
\begin{align}
M_{\rm ex}=\frac{1}{l}\biggl(\frac{n-1}{n-3}\biggl)^{(n-1)^2/4}\biggl(\frac{n-3}{2}\biggl)^{(n-1)/2}.
\end{align}

In this case, the mass function is given by 
\begin{align}
{\cal M}=\frac{R^{n-1}(1-Z)}{\{l^{2(n-2)/(n-1)}R^{4/(n-1)}-l^{2n/(n-1)}(1-Z)^{2/(n-1)}\}^{(n-1)/2}}
\end{align}
and hence
\begin{align}
\frac{\partial{\cal M}}{\partial R}=&\frac{R^{n-2}(1-Z)\{(n-3)l^{2(n-2)/(n-1)}R^{4/(n-1)}-(n-1)l^{2n/(n-1)}(1-Z)^{2/(n-1)}\}}{\{l^{2(n-2)/(n-1)}R^{4/(n-1)}-l^{2n/(n-1)}(1-Z)^{2/(n-1)}\}^{(n+1)/2}},\\
\frac{\partial{\cal M}}{\partial Z}=&-\frac{l^{2(n-2)/(n-1)}R^{(n^2-2n+5)/(n-1)}}{\{l^{2(n-2)/(n-1)}R^{4/(n-1)}-l^{2n/(n-1)}(1-Z)^{2/(n-1)}\}^{(n+1)/2}}.
\end{align}
Comparing this to Eqs.~(\ref{dMdZ}) and (\ref{dMdR}), we find that

\bea
\eta(R,Z)&=&2\phi_{,RR}Z+\frac{R^{n-2}(1-Z)\{(n-3)l^{2(n-2)/(n-1)}R^{4/(n-1)}-(n-1)l^{2n/(n-1)}(1-Z)^{2/(n-1)}\}}{\{l^{2(n-2)/(n-1)}R^{4/(n-1)}-l^{2n/(n-1)}(1-Z)^{2/(n-1)}\}^{(n+1)/2}}, \nonumber\\
\chi(R,Z)&=&\phi_{,R}-\frac{l^{2(n-2)/(n-1)}R^{(n^2-2n+5)/(n-1)}}{\{l^{2(n-2)/(n-1)}R^{4/(n-1)}-l^{2n/(n-1)}(1-Z)^{2/(n-1)}\}^{(n+1)/2}}.
\eea

\subsubsection{New nonsingular black hole}
Another physical nonsingular black hole is 
\begin{align}
f(R)=1+\frac{R^{n+1}}{2l^{n+2}M}\biggl(1- \sqrt{1+\frac{4l^{2n}M^2}{R^{2(n-1)}}}\biggl). \label{BD-type}
\end{align}
This metric resembles the vacuum solution in Einstein-Gauss-Bonnet gravity~\cite{Boulware1985,Wheeler1986}.
The metric reduces to de~Sitter and therefore the curvature is bounded for large $M$.

The mass-horizon relation is 
\begin{align}
M=&\frac{R_{\rm h}^{n+1}}{l^{n-2}(R_{\rm h}^4-l^4)}
\end{align}
and the global structure is similar to that of the Hayward black hole.
For this new black hole, the mass parameter for an extremal black hole is given by 
\begin{align}
M_{\rm ex}=\frac{n-3}{2l}\biggl(\frac{n+1}{n-3}\biggl)^{(n+1)/4}.
\end{align}

In this case, the mass function is given by 
\begin{align}
{\cal M}=\frac{R^{n+1}(1-Z)}{l^{n-2}R^4-l^{n+2}(1-Z)^2}
\end{align}
and hence
\begin{align}
\frac{\partial{\cal M}}{\partial R}=&\frac{(n-3)l^{n-2}R^{n+4}(1-Z)-(n+1)l^{n+2}R^{n}(1-Z)^3}{\{l^{n+2}(1-Z)^2-l^{n-2}R^4\}^2},\\
\frac{\partial{\cal M}}{\partial Z}=&-\frac{l^{n-2}R^{n+5}+l^{n+2}R^{n+1}(1-Z)^2}{\{l^{n+2}(1-Z)^2-l^{n-2}R^4\}^2}.
\end{align}
Comparing this to Eqs.~(\ref{dMdZ}) and (\ref{dMdR}), we see that
\bea
\eta(R,Z)&=&2\phi_{,RR}Z+\frac{(n-3)l^{n-2}R^{n+4}(1-Z)-(n+1)l^{n+2}R^{n}(1-Z)^3}{\{l^{n+2}(1-Z)^2-l^{n-2}R^4\}^2}, \nonumber\\
\chi(R,Z)&=&\phi_{,R}-\frac{l^{n-2}R^{n+5}+l^{n+2}R^{n+1}(1-Z)^2}{\{l^{n+2}(1-Z)^2-l^{n-2}R^4\}^2}.
\eea

\subsection{Designer Lovelock gravity}
We define {\it designer Lovelock (dL) gravity} by the dimensionally reduced action (\ref{eq:SphericalLovelockAction}) for spherically symmetric ELL gravity, but assume that all the ${\tilde \alpha}_{(p)}$ are potentially  nonzero for any value of $n$. In this case the action can no longer be lifted to a higher-dimensional ELL gravity since the corresponding Lovelock terms vanish identically for $p>n/2$. It does nonetheless provide us with an interesting 2D generalization of the spherical theory that can be interpreted in one of two ways: \\[5pt]
(i) the large coupling limit ${\alpha}_{(p)}=\infty$ for $p\ge [(n-1)/2]$,\\[5pt]
 or\\[5pt]
(ii) the large $n$ limit ($n\to \infty$).

Under the above assumption, the metric function $f(R)$ is determined just as in ELL gravity by
\begin{align}
\frac{l^{n-2}M}{(n-2)R^{n-1}}=& \sum_{p=0}^{\infty}{\tilde
\alpha}_{(p)}\biggl(\frac{1-f(R)}{R^2}\biggl)^p. \label{key-dL}
\end{align}
Since the right-hand side is an infinite series, it may be written as an analytic function by choosing ${\tilde
\alpha}_{(p)}$ appropriately:
\begin{align}
\frac{l^{n-2}M}{(n-2)R^{n-1}}=& \hw, \label{design}
\end{align}
where
\begin{align}
\nu:=&\frac{1-f(R)}{R^2}.
\end{align}

A given dL gravity action is therefore determined by a free analytic function $\hw$ that in turn determines the vacuum solution. 
The Hayward black hole (\ref{Hayward-BH-n}) is realized in dL gravity by choosing the coupling constants to give
\begin{align}
\hw=&\frac{\nu}{(n-2)(1-l^2\nu)} \nonumber \\
=&\frac{\nu}{n-2}\{1+l^2\nu+(l^2\nu)^2+\cdots\}.
\end{align}

Furthermore, with the following choice; 
\begin{align}
\hw=&\frac{\nu}{(n-2)\{1-(l^2\nu)^2\}} \nonumber \\
=&\frac{\nu}{n-2}\{1+(l^2\nu)^2+(l^2\nu)^{4}+\cdots\},
\end{align}
the new nonsingular black hole (\ref{BD-type}) is realized.
Although the metric resembles the vacuum solution in Einstein-Gauss-Bonnet gravity~\cite{Boulware1985,Wheeler1986}, the solution is realized  with only odd-order Lovelock terms in the action.

On the other hand, the Bardeen-type black hole (\ref{Bardeen-type}) is realized for 
\begin{align}
s(\nu)=\frac{\nu}{(n-2)\{1-l^{4/(n-1)}\nu^{2/(n-1)}\}^{(n-1)/2}}.
\end{align}
This cannot be realized as $s(\nu)=\sum_{p=0}^\infty {\tilde\alpha}_{(p)}\nu^p$ for any choice of ${\tilde\alpha}_{(p)}$. 
Hence, this nonsingular black hole is not realized in dL gravity and one needs to consider the full XL action (\ref{eq:lagrangianL}).

\section{Summary and future prospects}

We have presented a new class of gravity theories in two space-time dimensions.
The action contains three arbitrary functions and may provide a reasonable 2D effective theory for the spherically symmetric sector of a large class of higher-dimensional gravity theories.
Our actions are readily understood as extensions of spherically symmetric Einstein-Lanczos-Lovelock gravity. 
They share many, if not all, of the latter's desirable properties.

As shown in section~\ref{sec2}, the field equations are  second order, which implies  that the theories are ghost-free.
We have also identified the integrability condition for the theories to admit as first integral, a mass function that coincides with the generalized Misner-Sharp quasi-local mass in spherically symmetric ELL gravity~\cite{Maeda2011}.
The Hamiltonian analysis performed in section~\ref{sec:hamilton} showed that the super-Hamiltonian of the system is proportional to minus the spatial derivative of the mass function, and that, as a consequence, the on-shell mass function is a constant, both spatially and with respect to time.

As a consequence of the existence of the mass function, the system obeys Birkhoff's theorem.
In contrast to ELL gravity, the extensions admit a large class of static black holes as unique vacuum solutions that are nonsingular and have bounded curvature for arbitrarily large mass. 
In section~\ref{sec5}, we presented  examples of some physical nonsingular black-hole solutions. 
A subset of these nonsingular black holes are realized in what we call {\it designer Lovelock gravity}, which is understood as spherically symmetric ELL gravity with infinitely large coupling constants or large spacetime dimensions.

One natural and important question concerns which of the extended 2D theories that we have constructed can be obtained by imposing spherical symmetry in a fully covariant higher dimensional, higher curvature theory of gravity. 
In this context, a class of higher dimensional, higher curvature theories called ``quasi-topological gravity'' may play an important role~\cite{Oliva2011,Myers2010}.
Unlike ELL gravity, these theories are actually higher derivative theories whose 
field equations nonetheless become second order in spherically symmetric spacetimes. They therefore admit a mass function, and obey Birkhoff's theorem.
Although these theories do not admit nonsingular black holes, it is of great interest to understand the connection between our class of 2D dilaton gravity theories and dimensionally reduced quasi-topological gravity.
Indeed, it would be very valuable to be able to identify the most general class of higher-dimensional theories giving  second-order field equations in spherically symmetric spacetimes.

An important physical application of our new theories is the formation of a nonsingular black hole via gravitational collapse.
Quantum effects are expected to resolve the classical singularity in general relativity. While the notion of space and time itself may very well break down in the vicinity of the classically singularity, it is possible, and perhaps even likely, that the quantum effects could alter the conformal structure of the spacetime commonly thought to represent the formation and evaporation via Hawking radiation of black holes. This conformal structure lies at the heart of the information loss conundrum. In order to determine whether or not the absence of a singularity can potentially solve the information loss problem, it is necessary to have models that allow a quantitative study of the formation and evaporation of nonsingular black holes. In this context, our new 2D dilaton gravity may play an important role since the dynamics of quantum corrected nonsingular spacetimes can in principle be modeled by an effective theory of the form that we have presented.

\subsection*{Acknowledgments}
GK is grateful Jack Gegenberg, Viqar Husain, Jorma Louko, and Jon Ziprick for helpful conversations and comments on the manuscript.  
He also thanks Valeri Frolov for providing motivation to look for an action that could produce nice, nonsingular, spherically symmetric black-hole solutions. 
 HM thanks the Theoretical Physics group in University of Winnipeg for hospitality and support, where this work was started.
This work was funded in part by the Natural Sciences and Engineering Research Council of Canada.  Support was also provided by the Perimeter Institute for Theoretical Physics (funded by Industry Canada and the Province of Ontario Ministry of Research and Innovation).
This work has also been funded by the Fondecyt Grant No. 3140123. The Centro de Estudios Cient\'{\i}ficos (CECs) is funded by the Chilean Government through the Centers of Excellence Base Financing Program of Conicyt.

\appendix

\section{Details of derivations}
\label{appendix}
\subsection{Lagrangian density}
\label{app:Lagrangian}
Here we present how to derive the Lagrangian density (\ref{eq:IDeltaIFinal}).
The Lagrangian density for the action (\ref{def-ILI}) in terms of ADM variables is given by 
\begin{align}
\label{eq:DeltaL}
{} {\cal L}^{(I)}_{\rm L} &= \frac{1}{l^{n-2}}\Ialpha(R)W_{(I)}(Z)\Lambda N
\biggl\{R_{,t}Z_{,t}g^{00}
     + \left(R_{,x} Z_{,t}
                     +R_{,t} Z_{,x} \right)g^{01} 
        +R_{,x} Z_{,x} g^{11}\biggl\}\nonumber\\
&=  \frac{1}{l^{n-2}}\Ialpha(R)W_{(I)}(Z)\Lambda N\biggl\{
     -\frac{R_{,t}Z_{,t}}{N^2}+ \frac{N_r}{N^2}\left(R_{,x}Z_{,t}+R_{,t}Z_{,x}\right)
  +R_{,x}Z_{,x}\left(\frac{1}{\Lambda^2}-\frac{N_r^2}{N^2}\right)\biggl\}\nonumber\\
 &= -\frac{1}{l^{n-2}}\Ialpha(R)W_{(I)}(Z)\Lambda N \left(R_{,u}Z_{,u}-\frac{1}{\Lambda^2}R_{,x}Z_{,x}\right).
\end{align}

Note that in this appendix we drop the sum over $I$ for simplicity.  The first term in the above is the interesting one since we need to eliminate the second time derivatives in $Z_{,u}$ in order to put the action into the Hamiltonian form. Inspired by the appendix in~\cite{Taves2013}, we write
\begin{align}
\label{eq:DeltaL2}
{} {\cal L}^{(I)}_{\rm L}
 &= -\frac{1}{l^{n-2}}\Ialpha(R)W_{(I)}(Z)\Lambda N
   \left\{R_{,u}\left(\frac{1}{N}Z_{,t}-\frac{N_r}{N}Z_{,x}\right)-\frac{R_{,x}}{\Lambda^2}{Z_{,x}}\right\}\nonumber\\
  &= {}{\cal L}^{(I)}_{\rm L1}+{}{\cal L}^{(I)}_{\rm L2}+{}{\cal L}^{(I)}_{\rm L3},
\end{align}
where
\bea
{}  {\cal L}^{(I)}_{\rm L1}&:=&-\frac{1}{l^{n-2}}\Ialpha(R) \Lambda W_{(I)}(Z)  R_{,u} Z_{,t},
\nonumber\\
{}  {\cal L}^{(I)}_{\rm L2}&:=&\frac{1}{l^{n-2}}\Ialpha(R) {N_r}\Lambda W_{(I)}(Z) R_{,u}Z_{,x},
\nonumber\\
{}  {\cal L}^{(I)}_{\rm L3}&:=&\frac{1}{l^{n-2}}\frac{N}{\Lambda}\Ialpha(R){R_{,x}} W_{(I)}(Z) {Z_{,x}}.
\eea

We note that
\be
\delta{Z} = -2R_{,u}\delta R_{,u} + \delta b,
\ee
where we have defined for convenience
\be
b:= \frac{{R_{,x}}^2}{\Lambda^2}.
\ee
We now 
assume that $W_{(I)}(Z)(=W_{(I)}(-{R_{,u}}^2 + b))$ has a Taylor expansion in $Z$ and hence in ${R_{,u}}^2$, so that
\be
W_{(I)}(Z) = \sum_{m} \Hb {R_{,u}}^{2m},
\ee
where
\be
\Hb := \frac{(-1)^m}{m ! }\left. \frac{\D^m W_{(I)}}{\D Z^m}\right|_{Z=b}.
\ee
Then, we have
\bea
W_{(I)}(Z)R_{,u}\delta Z &=& W_{(I)}(Z)R_{,u}(-2R_{,u}\delta R_{,u}+\delta b)\nonumber\\
    &=&-2 \sum_{m} \Hb {R_{,u}}^{2m+2}\delta R_{,u} + W_{(I)}R_{,u}\delta b\nonumber\\
   &=& -2\sum_{m} \Hb \frac{\delta ({R_{,u}}^{2m+3})}{2m+3}  +W_{(I)}R_{,u}\delta b.
\eea
The first term can now be integrated by parts term by term. This gives, up to total derivatives,
\begin{align}
{} {\cal L}^{(I)}_{\rm L1}:=&-\frac{1}{l^{n-2}}\Ialpha(R)W_{(I)}(Z)  \Lambda R_{,u} Z_{,t}\nonumber\\
  =& -\frac{1}{l^{n-2}}\Ialpha(R)\Lambda\left\{- 2\sum_{m} \Hb \frac{({R_{,u}}^{2m+3})_{,t}}{2m+3} +W_{(I)}(Z)R_{,u} b_{,t}     \right\}\nonumber\\
=&\frac{1}{l^{n-2}}\left\{- 2\sum_{m}\left(\Ialpha(R)\Lambda \Hb \right)_{,t}\frac{{R_{,u}}^{2m+3}}{2m+3}  -\Ialpha(R)\Lambda W_{(I)}(Z)R_{,u} b_{,t}     \right\}.
\end{align}
Similarly we obtain
\begin{align}
{} {\cal L}^{(I)}_{\rm L2}
=&\frac{1}{l^{n-2}}\left\{2\sum_{m}\left(N_r\Ialpha(R) \Lambda \Hb \right)_{,x}\frac{{R_{,u}}^{2m+3}}{2m+3}  +N_r\Ialpha(R) \Lambda W_{(I)}(Z)R_{,u} b_{,x}     \right\}.
\end{align}
Finally, ${}{\cal L}^{(I)}_{\rm L3}$ is straightforward because of
\be
W_{(I)}(Z)Z_{,x}= (X_{(I)}(Z))_{,x},
\ee
where 
\be
W_{(I)}(Z) = X_{(I)}(Z)_{,Z}.
\ee
Thus ${}{\cal L}^{(I)}_{\rm L3}$ is computed as
\bea
{}{\cal L}^{(I)}_{\rm L3} &=& \frac{1}{l^{n-2}}\frac{N}{\Lambda}\Ialpha(R)W_{(I)}(Z){R_{,x}}{Z_{,x}}\nonumber\\
  &=& \frac{1}{l^{n-2}}\frac{N}{\Lambda}\Ialpha(R){R_{,x}}{X_{(I)}(Z)_{,x}}\nonumber\\
  &=& -\left(\frac{1}{l^{n-2}}\frac{N}{\Lambda}\Ialpha(R){R_{,x}}\right)_{,x}{X_{(I)}(Z)}
\eea
up to boundary terms.

Putting it all together, we have
\bea
{}{\cal L}^{(I)}_{L}
 &=& \frac{1}{l^{n-2}}\left\{- 2\sum_{m}\left(\Ialpha(R) \Lambda \Hb \right)_{,t}\frac{{R_{,u}}^{2m+3}}{2m+3}  -\Ialpha(R)\Lambda  W_{(I)}(Z)R_{,u} b_{,t}  \right\} \nonumber\\
  & & +\frac{1}{l^{n-2}}\left\{ 2\sum_{m}\left(N_r\Ialpha(R)\Lambda \Hb \right)_{,x}\frac{{R_{,u}}^{2m+3}}{2m+3} +N_r\Ialpha(R)\Lambda W_{(I)}(Z)R_{,u} b_{,x}     \right\}
\nonumber\\
& & -\left(\frac{1}{l^{n-2}}\frac{N}{\Lambda}\Ialpha(R){R_{,x}}\right)_{,x}{X_{(I)}(Z)}.
\label{eq:TotalDeltaI}
\eea

We will concentrate on the first line of the above, since it has the time derivatives of $\Lambda$. We will need
\be
b_{,t} = 2\frac{R_{,x}}{\Lambda^2} R_{,tx} - \frac{2{R_{,x}}^2}{\Lambda^3}\Lambda_{,t}.
\ee
Thus we have
\bea
l^{n-2}{}{\cal L}^{(I)}_{\rm L1} &=& - 2\sum_{m}\left(\Ialpha(R) \Lambda \Hb \right)_{,t}\frac{{R_{,u}}^{2m+3}}{2m+3} -\Ialpha(R)\Lambda  W_{(I)}(Z)R_{,u} b_{,t} \nonumber\\
   &=&  - 2\sum_{m}\biggl\{\Ialpha(R)_{,t} \Lambda \Hb
 + \Ialpha(R) \Lambda_{,t} \Hb \nonumber\\
& &   + 
    \Ialpha(R) \Lambda \frac{\D \Hb }{\D b} \left( 2\frac{R_{,x}}{\Lambda^2} R_{,xt} - \frac{2{R_{,x}}^2}{\Lambda^3}\Lambda_{,t}  \right) \biggl\}\frac{{R_{,u}}^{2m+3}}{2m+3} \nonumber \\
&&    -\Ialpha(R)\Lambda  W_{(I)}(Z)R_{,u} \left(2\frac{R_{,x}}{\Lambda^2} R_{,xt} - \frac{2{R_{,x}}^2}{\Lambda^3}\Lambda_{,t}\right).  
\eea
Collecting time derivatives of $\Lambda$, we get
\begin{align}
l^{n-2}{}{\cal L}^{(I)}_{\rm L1} =& - 2\sum_{m}\left\{
  \Ialpha(R) \Hb \Lambda_{,t} + 
    \Ialpha(R) \Lambda\frac{\D \Hb}{\D b} \left( - \frac{2{R_{,x}}^2}{\Lambda^3}\Lambda_{,t}  \right) \right\}\frac{{R_{,u}}^{2m+3}}{2m+3}  \nonumber\\ 
 & -\Ialpha(R)\Lambda  W_{(I)}(Z)R_{,u} \left( - \frac{2{R_{,x}}^2}{\Lambda^3}\Lambda_{,t}\right) \nonumber \\
&- 2\sum_{m}\left\{\Ialpha(R)_{,t} \Lambda \Hb
 +
    \Ialpha(R) \Lambda\frac{\D \Hb}{\D b} \left( 2\frac{R_{,x}}{\Lambda^2}R_{,xt}  \right) \right\}\frac{{R_{,u}}^{2m+3}}{2m+3} \nonumber\\
 & 
    -\Ialpha(R)\Lambda  W_{(I)}(Z)R_{,u} \left(2\frac{R_{,x}}{\Lambda^2} R_{,xt} \right) 
\nonumber\\
 =&\left[
 - 2\sum_{m}\left\{
 \Ialpha(R) \Hb  + 
    \Ialpha(R) \Lambda\frac{\D \Hb }{\D b} \left( - \frac{2{R_{,x}}^2}{\Lambda^3} \right) \right\}\right.\frac{{R_{,u}}^{2m+3}}{2m+3}\nonumber\\
 &\left. 
    -\Ialpha(R)\Lambda  W_{(I)}(Z)R_{,u} \left( - \frac{2{R_{,x}}^2}{\Lambda^3}\right) \right]\Lambda_{,t} \nonumber\\ 
 & - 2\sum_{m}\left\{\Ialpha(R)_{,t} \Lambda \Hb
 +
    \Ialpha(R) \Lambda \frac{\D \Hb }{\D b} \left( 2\frac{R_{,x}}{\Lambda^2}R_{,xt}  \right) \right\}  \frac{{R_{,u}}^{2m+3}}{2m+3} \nonumber\\
 &-\Ialpha(R)\Lambda  W_{(I)}(Z)R_{,u} \left(2\frac{R_{,x}}{\Lambda^2}R_{,xt} \right) 
\nonumber\\
  =&l^{n-2} {} P_\Lambda^{(I)} \Lambda_{,t}  -\Ialpha(R)\Lambda  W_{(I)}(Z)R_{,u} \left(2\frac{R_{,x}}{\Lambda^2} R_{,xt} \right) \nonumber\\
 & - 2\sum_{m}\left\{\Ialpha(R)_{,t} \Lambda \Hb
 +
    \Ialpha(R)\Lambda \frac{\D \Hb}{\D b} \left( 2\frac{R_{,x}}{\Lambda^2}R_{,xt}  \right) \right\}\frac{{R_{,u}}^{2m+3}}{2m+3},
\label{eq:DeltaI1}
\end{align}
where
\begin{align}
{} P^{(I)}_\Lambda:=&
\frac{1}{l^{n-2}}\Bigg[- 2\sum_{m}\Ialpha(R) \left\{\Hb-2 \frac{\D \Hb}{\D b}\left(\frac{{R_{,x}}^2}{\Lambda^2}\right)\right\}\frac{{R_{,u}}^{2m+3}}{2m+3} \nonumber \\
&  +2\Ialpha(R)  W_{(I)}(Z)R_{,u}\frac{{R_{,x}}^2}{\Lambda^2}   \Bigg] \label{eq:DeltaPLambda}
\end{align}
is the contribution to the conjugate momentum of $\Lambda$ from ${}{\cal L}^{(I)}_{\rm L}$.


We now repeat this for the second line of (\ref{eq:DeltaI1}).
Noting that it is exactly the same form as the first line, we obtain
\bea
l^{n-2}{}{\cal L}^{(I)}_{\rm L2} &=&  2\sum_{m=0}\left(N_r\Ialpha(R)\Lambda \Hb \right)_{,x}\frac{{R_{,u}}^{2m+3}}{2m+3}  +N_r\Ialpha(R)\Lambda {W_{(I)}}(Z)R_{,u} b_{,x} 
\nonumber\\
  &=&   2N_r\sum_{m=0}\left\{\frac{N_{r,x}}{N_r}\Ialpha(R) \Lambda \Hb +\Ialpha(R)_{,x} \Lambda \Hb
 + \Ialpha(R) \Lambda_{,x} \Hb \right. \nonumber\\
& &\left. +
    \Ialpha(R) \Lambda\frac{\D \Hb }{\D b} \left( 2\frac{R_{,x}}{\Lambda^2}{R}_{,xx} - \frac{2{R_{,x}}^2}{\Lambda^3}\Lambda_{,x}  \right) \right\}  \frac{{R_{,u}}^{2m+3}}{2m+3}\nonumber\\
& &+N_r\Ialpha(R)\Lambda  W_{(I)}(Z)R_{,u} \left(2\frac{R_{,x}}{\Lambda^2}{R}_{,xx} - \frac{2{R_{,x}}^2}{\Lambda^3}\Lambda_{,x}\right). 
\eea
Collecting terms in $\Lambda_{,x}$, we get
\bea
l^{n-2}{}{\cal L}^{(I)}_{\rm L2} &=&
 N_r \left[
  2\sum_{m}\left\{
  \Ialpha(R) \Hb  + 
    \Ialpha(R) \Lambda \frac{\D \Hb}{\D b} \left( - \frac{2{R_{,x}}^2}{\Lambda^3} \right) \right\}\right.\frac{{R_{,u}}^{2m+3}}{2m+3} \nonumber\\
& &\left.      +\Ialpha(R)\Lambda  W_{(I)}(Z)R_{,u} \left( - \frac{2{R_{,x}}^2}{\Lambda^3}\right) \right]\Lambda_{,x}  +N_r\Ialpha(R)\Lambda  W_{(I)}(Z)R_{,u} \left(2\frac{R_{,x}}{\Lambda^2}{R}_{,xx}\right) \nonumber\\ 
& & + 2N_r\sum_{m}\left\{\frac{N_{r,x}}{N_r}\Ialpha(R) \Lambda \Hb +\Ialpha(R)_{,x} \Lambda \Hb
 \right.\nonumber\\
& &\left. +
    \Ialpha(R) \Lambda\frac{\D \Hb}{\D b} \left( 2\frac{R_{,x}}{\Lambda^2}{R}_{,xx}   \right) \right\}\frac{{R_{,u}}^{2m+3}}{2m+3} \nonumber\\
 &=& -N_r {} P^{(I)}_\Lambda \Lambda_{,x} l^{n-2}  +N_r\Ialpha(R)\Lambda  W_{(I)}(Z)R_{,u} \left(2\frac{R_{,x}}{\Lambda^2}{R}_{,xx}\right) \nonumber\\
  & &+ 2N_r\sum_{m}\left\{\frac{N_{r,x}}{N_r}\Ialpha(R) \Lambda \Hb +\Ialpha(R)_{,x} \Lambda \Hb
 \right. \nonumber\\
& & \left. +
    \Ialpha(R) \Lambda\frac{\D \Hb }{\D b} \left( 2\frac{R_{,x}}{\Lambda^2}{R}_{,xx}   \right) \right\}\frac{{R_{,u}}^{2m+3}}{2m+3}. \label{eq:DeltaI2}
\eea

The third line of (\ref{eq:TotalDeltaI}) can be integrated by parts to give
\be
{}{\cal L}^{(I)}_{\rm L3}= \frac{1}{l^{n-2}}\frac{N}{\Lambda}\Ialpha(R){R_{,x}} {W_{(I)}(Z)}Z_{,x}.
\label{eq:DeltaI3}
\ee

Putting together (\ref{eq:DeltaI1}), (\ref{eq:DeltaI2}), and (\ref{eq:DeltaI3}) and replacing $R_{,t}$ by $NR_{,u}+N_rR_{,x}$, we get
\bea
l^{n-2}{}{\cal L}^{(I)}_{\rm L} &=&l^{n-2}\biggl({} P^{(I)}_\Lambda \Lambda_{,t} - N_r{} P^{(I)}_\Lambda \Lambda_{,x}
    -{N_r}_{,x}{} P^{(I)}_\Lambda \Lambda\biggl)\nonumber\\
 & & -2(NR_{,u})_{,x}\left(\Ialpha(R)W_{(I)}(Z)R_{,u} \frac{R_{,x}}{\Lambda} + 
      2\Ialpha(R) \frac{R_{,x}}{\Lambda} \sum_m \frac{\D \Hb }{\D b}\frac{{R_{,u}}^{2m+3}}{2m+3}
      \right)\nonumber\\
 & &  -2N\Ialpha{}_{,R} \Lambda R_{,u} \sum_m \Hb \frac{{R_{,u}}^{2m+3}}{2m+3}
    + \frac{N}{\Lambda}\Ialpha(R){R_{,x}}{W_{(I)}(Z)}Z_{,x}  \nonumber\\
 &=&l^{n-2}\biggl({} P^{(I)}_\Lambda \Lambda_{,t} + N_r \Lambda {P^{(I)}_\Lambda}_{,x} 
    -(N_r{} P^{(I)}_\Lambda \Lambda)_{,x}\biggl)\nonumber\\
& &  +2NR_{,u}\left( \Ialpha(R)W(Z)R_{,u} \frac{R_{,x}}{\Lambda} + 
      2\Ialpha(R) \frac{R_{,x}}{\Lambda} \sum_m \frac{\D \Hb }{\D b}\frac{{R_{,u}}^{2m+3}}{2m+3}\right)
      _{,x}\nonumber\\
 & &  -2N\Ialpha{}_{,R} \Lambda R_{,u} \sum_m \Hb \frac{{R_{,u}}^{2m+3}}{2m+3}
  +\frac{N}{\Lambda}\Ialpha(R){R_{,x}}{W_{(I)}(Z)}Z_{,x}.
\label{eq:DeltaIFinal}
\eea
The first term will cancel with the corresponding Liouville term when constructing the Hamiltonian density. The second term gives the expected contribution to the diffeomorphism constraint. The next two  terms give  non-trivial contributions to the Hamiltonian constraint.

\subsection{Hamiltonian density} 
\label{app:Hamiltonian}
Here we present how to derive the Hamiltonian density (\ref{eq:IFinalH}).
 The momentum conjugate to $\Lambda$ is
\be
P_\Lambda = -\frac{2}{l^{n-2}} \phi_{,R}R_{,u} +\sum_{I} P^{(I)}_\Lambda,
\label{eq:FullPLambda2}
\ee
where the second term is given in (\ref{eq:DeltaPLambda}). The total Hamiltonian density is then
\bea
{\cal H}_{\rm XL} &=&  P_{\Lambda}\Lambda_{,t} + P_R R_{,t} - {\cal L}_{\rm XL}\nonumber\\
 &=& N{\cal H} + N_r{\cal H}_r,
\eea
where now
\bea
{\cal H}&:=& P_R R_{,u}   + \frac{1}{l^{n-2}}\sum_{I}\Bigg\{2\left(\frac{\phi_{,x}}{\Lambda}\right)_{,x}-{\Lambda}\HRZ \nonumber\\
  & & -2R_{,u}\left( \Ialpha(R)W(Z)R_{,u} \frac{R_{,x}}{\Lambda} + 
      2\Ialpha(R) \frac{R_{,x}}{\Lambda} \sum_m \frac{\D \Hb}{\D b}\frac{{R_{,u}}^{2m+3}}{2m+3}
      \right)_{,x}\nonumber\\
 & &  +2\Ialpha{}_{,R} \Lambda R_{,u} \sum_m \Hb \frac{{R_{,u}}^{2m+3}}{2m+3}
    - \frac{R_{,x}}{\Lambda}\Ialpha(R){W(Z)}Z_{,x} \Bigg\}, \\
{\cal H}_r &:=& P_R R_{,x} - {P_\Lambda}_{,x}\Lambda. 
\eea
In the above $R_{,u}$ is an implicit function of $\Lambda$, $P_\Lambda$, and $R$ given by (\ref{eq:FullPLambda}).

As before we want to eliminate $P_R$ completely from the Hamiltonian constraint in the hopes of finding a suitable mass function, so we define
\bea
\tilde{H} &:=& \frac{R_{,x}}{ \Lambda}{\cal H} -\frac{R_{,u}}{ \Lambda}{\cal H}_r \nonumber\\
    &=& R_{,u}{P_\Lambda}_{,x} + \frac{1}{l^{n-2}}\sum_{I} \Bigg\{\frac{2R_{,x}}{\Lambda}
  \left(\frac{\phi_{,x}}{\Lambda}\right)_{,x}  - \HRZ R_{,x}\nonumber\\
  & & -2R_{,u}\frac{R_{,x}}{\Lambda}\left( \Ialpha(R)W_{(I)}(Z)R_{,u} \frac{R_{,x}}{\Lambda} + 
      2\Ialpha(R) \frac{R_{,x}}{\Lambda} \sum_m \frac{\D \Hb }{\D b}\frac{{R_{,u}}^{2m+3}}{2m+3}\right)_{,x}\nonumber\\
 & &  +2\Ialpha{}_{,R} R_{,x} R_{,u} \sum_m \Hb \frac{{R_{,u}}^{2m+3}}{2m+3}  
  - \left(\frac{R_{,x}}{\Lambda}\right)^2\Ialpha(R){W(Z)}Z_{,x} \Bigg\}.
\label{eq:TildeH1}
\eea
We need to express the first term in terms of $R_{,u}$:
\begin{align}
R_{,u}{P_\Lambda}_{,x}=& -\frac{2}{l^{n-2}}R_{,u}\sum_{I}\Bigg\{  \phi_{,R}R_{,u} \nonumber\\
&+  \sum_{m}\Ialpha(R) \left(\Hb -2 \frac{\D \Hb }{\D b}\frac{{R_{,x}}^2}{\Lambda^2}\right)\frac{{R_{,u}}^{2m+3}}{2m+3}
  -\Ialpha(R)  W_{(I)}(Z)R_{,u}\frac{{R_{,x}}^2}{\Lambda^2}   \Bigg\}_{,x}\nonumber\\
 =&  \frac{1}{l^{n-2}}\sum_{I}\left[-\frac{1}{\phi_{,R}}  \left((\phi_{,R}R_{,u})^2 \right)_{,x} \right. \nonumber\\
-& \left.  2R_{,u}\left\{\sum_{m}\Ialpha(R) \left(\Hb -2 \frac{\D \Hb }{\D b}\frac{{R_{,x}}^2}{\Lambda^2}\right)\frac{{R_{,u}}^{2m+3}}{2m+3}
  -\Ialpha(R)  W_{(I)}(Z)R_{,u}\frac{{R_{,x}}^2}{\Lambda^2}   \right\}_{,x}\right]. \label{eq:RP}
\end{align}
Putting (\ref{eq:RP}) into (\ref{eq:TildeH1}) yields
\begin{align}
l^{n-2}\tilde{H} :=& \frac{1}{\phi_{,R}}\left\{ - \left((\phi_{,R}R_{,u})^2 \right)_{,x} 
 +\frac{2\phi_{,R}R_{,x}}{\Lambda}\left(\frac{\phi_{,x}}{\Lambda}\right)_{,x}\right\}  -\HRZ R_{,x} \nonumber \\
&    - \left(\frac{R_{,x}}{\Lambda}\right)^2\sum_{I}\Ialpha(R){W_{(I)}(Z)}Z_{,x} +\sum_{I}\Ialpha W_{(I)} {R_{,u}}^2 \left(\frac{{R_{,x}}^2}{\Lambda^2} \right)_{,x}\nonumber\\
 & + \sum_{I}2\Ialpha R_{,u}\left(\frac{{R_{,x}}^2}{\Lambda^2} \right)_{,x}\sum_m \frac{\D \Hb }{\D b}\frac{{R_{,u}}^{2m+3}}{2m+3}
 -\sum_{I}2\Ialpha R_{,u}\left(\sum_m \Hb \frac{{R_{,u}}^{2m+3}}{2m+3} \right)_{,x} \nonumber\\
=&\frac{1}{\phi_{,R}}({\phi_{,R}}^2 Z)_{,x}  - \HRZ R_{,x}  - \left(\frac{R_{,x}}{\Lambda}\right)^2\sum_{I}\Ialpha(R){W_{(I)}(Z)}Z_{,x} 
\nonumber\\
 &+\sum_{I}\Ialpha W {R_{,u}}^2 \left(\frac{{R_{,x}}^2}{\Lambda^2} \right)_{,x} + \sum_{I}2\Ialpha R_{,u}\left(\frac{{R_{,x}}^2}{\Lambda^2} \right)_{,x}\sum_m \frac{\D \Hb }{\D b}\frac{{R_{,u}}^{2m+3}}{2m+3} \nonumber \\
& -\sum_{I}2\Ialpha R_{,u}\left(\sum_m \Hb \frac{{R_{,u}}^{2m+3}}{2m+3} \right)_{,x}.
\label{eq:TildeH2}
\end{align}
Expanding the derivative in the last term gives
\begin{align}
 -2\Ialpha R_{,u}&\left(\sum_m \Hb \frac{{R_{,u}}^{2m+3}}{2m+3} \right)_{,x}
 \nonumber\\
=& -2\Ialpha R_{,u}\sum_m \frac{\D \Hb }{\D b}\frac{{R_{,u}}^{2m+3}}{2m+3} b_{,x} -2\Ialpha R_{,u}\sum_m \Hb {R_{,u}}^{2m+2} (R_{,u})_{,x}\nonumber\\
 =&  -2\Ialpha R_{,u}\left(\frac{{R_{,x}}^2}{\Lambda^2} \right)_{,x}\sum_m \frac{\D \Hb}{\D b}\frac{{R_{,u}}^{2m+3}}{2m+3}  -2\Ialpha {R_{,u}}^3(R_{,u})_{,x}\sum_m \Hb {{R_{,u}}^{2m}} \nonumber\\
=& -2\Ialpha R_{,u}\left(\frac{{R_{,x}}^2}{\Lambda^2} \right)_{,x}\sum_m \frac{\D \Hb}{\D b}\frac{{R_{,u}}^{2m+3}}{2m+3}  -\Ialpha W_{(I)}{R_{,u}}^2 ({R_{,u}}^2)_{,x} .
\end{align}
Putting the above into (\ref{eq:TildeH2}) gives, quite miraculously,
\bea
\tilde{\cal H} &=& \frac{1}{\phi_{,R}}({\phi_{,R}}^2Z)_{,x}  - \HRZ R_{,x} \nonumber\\
&& +\sum_{I}\left\{ \Ialpha W_{(I)}(Z) {R_{,u}}^2 Z_{,x}  - \left(\frac{R_{,x}}{\Lambda}\right)^2\Ialpha(R){W_{(I)}(Z)}Z_{,x}\right\}\nonumber\\
 &=& \biggl(2\phi_{,RR}Z- \HRZ \biggl) R_{,x} + \biggl(\phi_{,R}-\chi(R,Z) Z\biggl) Z_{,x},
\label{eq:FinalH}
\eea
where $\chi(R,Z)$ is defined in (\ref{eq:chi}).  

\bibliography{NonSingularPaper2015GK}

\begin{thebibliography}{10}

\bibitem{Bojowald2006}
M.~Bojowald and A.~Skirzewski.
\newblock Effective equations of motion for quantum systems.
\newblock {\em Reviews of Mathematical Physics}, 18:713, 2006.

\bibitem{Chacon-Acosta2012}
G~Chac/'on-Acosta and H.H. Hern\'andez.
\newblock Effective quantum equations for the semiclassical description of the
  hydrogen atom.
\newblock {\em Arxiv preprint arXiv:1110.3337}, 2011.

\bibitem{Das2013}
Saurya Das.
\newblock Quantum raychaurduri equation.
\newblock {\em Physical Review D}, 89:084068, 2013.

\bibitem{Peltola2009a}
A.~Peltola and G.~Kunstatter.
\newblock {A Complete Single-Horizon Quantum Corrected Black Hole Spacetime}.
\newblock {\em Physical Review D}, 79:061501, 2009.

\bibitem{Peltola2009b}
A.~Peltola and G.~Kunstatter.
\newblock {Effective Polymer Dynamics of D-Dimensional Black Hole Interiors}.
\newblock {\em Physical Review D}, 80(1):044031, 1966.

\bibitem{Sakharov1966}
Ad~Sakharov.
\newblock {The Initial Stage of an Expanding Universe and the Appearance of a
  Nonuniform Distribution of Matter}.
\newblock {\em Soviet Physics JETP}, 22(1):241, 1966.

\bibitem{Bardeen1968}
J.~Bardeen.
\newblock {Non-Singular General-Relativistic Gravitational Collapse}.
\newblock {\em Presented at GR5, Tiblisi, U.S.S.R., and published in the
  conference proceedings in the U.S.S.R.}, 1968.

\bibitem{Poisson1988}
Eric Poisson and Werner Israel.
\newblock {Structure of the Black Hole Nucleus}.
\newblock {\em Classical Quantum Gravity}, 5:201--205, 1988.

\bibitem{Dymnikova2005}
Irina Dymnikova and Evgeny Galaktionov.
\newblock {Stability of a Vacuum Non-singular Black Hole}.
\newblock {\em Classical and Quantum Gravity}, 22(12):2331--2357, June 2005.

\bibitem{Ayon-Beato1999}
Eloy Ay\'{o}n-Beato and Alberto Garcı́a.
\newblock {Regular Black Hole in General Relativity Coupled to Nonlinear
  Electrodynamics}.
\newblock {\em Physical Review Letters}, 80(23):5056--5059, 1998.

\bibitem{Garcia99}
Eloy Ayon-Beato and Alberto Garcia.
\newblock New regular black hole solutin from nonlinear electrodynamics.
\newblock {\em Phys. Lett.}, B464:25, 1999.

\bibitem{Ayon-Beato1999a}
Eloy Ay\'{o}n-Beato.
\newblock {Non–Singular Charged Black Hole Solution for Non–Linear Source}.
\newblock {\em General Relativity and Gravitation}, 31(5):629--633, 1999.

\bibitem{Ayon-Beato2000}
Eloy Ayon-Beato and Alberto Garcia.
\newblock {The Bardeen model as a nonlinear magnetic monopole}.
\newblock {\em Phys. Lett.}, B493:149--152, 2000.

\bibitem{Ayon-Beato2005}
Eloy Ayon-Beato and Alberto Garcia.
\newblock {Four parametric regular black hole solution}.
\newblock {\em Gen. Rel. Grav.}, 37:635, 2005.

\bibitem{Grumiller2002}
Daniel Grumiller, W.~Kummer, and D.~V. Vassilevich.
\newblock {Dilaton Gravity in Two Dimensions}.
\newblock {\em Physics Reports}, 369(4):327--430, 2002.

\bibitem{Ziprick2010}
Jonathan Ziprick and Gabor Kunstatter.
\newblock {Quantum Corrected Spherical Collapse: A Phenomenological Framework}.
\newblock {\em Physical Review D}, 82(4):044031, 2010.

\bibitem{Ziprick2009}
Jonathan Ziprick.
\newblock {\em {Singularity Resolution and Dynamical Black Holes}}.
\newblock Msc, Manitoba, 2009.

\bibitem{Taves2014}
Tim Taves and Gabor Kunstatter.
\newblock {Modelling the Evaporation of Non-singular Black Holes}.
\newblock {\em Physical Review D}, 90(12):124062, 2014.

\bibitem{Grumiller2003}
Daniel Grumiller.
\newblock {Deformations of the Schwarzschild Black Hole}.
\newblock {\em arXiv preprint gr-qc/0311011}, 2003.

\bibitem{Grumiller2004}
Daniel Grumiller.
\newblock {Long Time Black Hole Evaporation with Bounded Hawking Flux}.
\newblock {\em Journal of Cosmology and Astroparticle Physics}, 2004(05):5,
  2004.

\bibitem{Hertog2005}
T.~Hertog and G.T. Horowitz.
\newblock Designer gravity and field theory effective potentials.
\newblock {\em Physical Review Letters}, 94:221301, 2005.

\bibitem{Maeda2008}
Hideki Maeda and Masato Nozawa.
\newblock {Generalized Misner-Sharp Quasi-local Mass in Einstein-Gauss-Bonnet
  Gravity}.
\newblock {\em Physical Review D}, 77(6):064031, 2008.

\bibitem{Palais1979}
R.~S. Palais.
\newblock {The Principle of Symmetric Criticality}.
\newblock {\em Communications in Mathematical Physics}, 69(1):19--30, 1979.

\bibitem{Fels2002}
Mark~E. Fels and Charles~G. Torre.
\newblock {The Principle of Symmetric Criticality in General Relativity}.
\newblock {\em Classical and Quantum Gravity}, 19(4):641--675, 2002.

\bibitem{Deser2003a}
S.~Deser and Bayram Tekin.
\newblock {Shortcuts to High Symmetry Solutions in Gravitational Theories}.
\newblock {\em Classical and Quantum Gravity}, 20(22):4877--4883, 2003.

\bibitem{Misner-Sharp}
C.~W. Misner and D.~H. Sharp.
\newblock {Relativistic Equations for Adiabatic, Spherically Symmetric
  Gravitational Collapse}.
\newblock {\em Phys. Rev.}, 136:B571, 1964.

\bibitem{Lanczos1938}
Cornelius Lanczos.
\newblock {A Remarkable Property of the Riemann-Christoffel Tensor in Four
  Dimensions}.
\newblock {\em The Annals of Mathematics}, 39(4):842, 1938.

\bibitem{Lovelock1971}
David Lovelock.
\newblock {The Einstein Tensor and Its Generalizations}.
\newblock {\em Journal of Mathematical Physics}, 12(3):498--501, 1971.

\bibitem{lovelockreview}
C.~Garraffo and G.~Giribet.
\newblock {The Lovelock Black Holes}.
\newblock {\em Phys. Lett. A}, 23:1801, 2008.

\bibitem{lovelockreview2}
C.~Charmousis.
\newblock {Higher Order Gravity Theories and Their Black Hole Solutions}.
\newblock {\em Physics Lecture Notes}, 2009.

\bibitem{Kunstatter2012}
Gabor Kunstatter, Tim Taves, and Hideki Maeda.
\newblock {Geometrodynamics of Spherically Symmetric Lovelock Gravity}.
\newblock {\em Classical and Quantum Gravity}, 29(9):092001, 2012.

\bibitem{Kunstatter2013}
Gabor Kunstatter, Hideki Maeda, and Tim Taves.
\newblock {Hamiltonian Dynamics of Lovelock Black Holes with Spherical
  Symmetry}.
\newblock {\em Classical and Quantum Gravity}, 30(6):065002, 2013.

\bibitem{Taves2013}
Tim Taves.
\newblock {\em {Black Hole Formation in Lovelock Gravity}}.
\newblock Phd, University of Manitoba, 2013.

\bibitem{Maeda2011}
Hideki Maeda, Steven Willison, and Sourya Ray.
\newblock {Lovelock Black Holes with Maximally Symmetric Horizons}.
\newblock {\em Classical and Quantum Gravity}, 28(16):165005, 2011.

\bibitem{zegers2005}
R.~Zegers.
\newblock {Birkhoff’s Theorem in Lovelock Gravity}.
\newblock {\em J. Math. Phys.}, 46(7):072502, 2005.

\bibitem{Wheeler1986}
James~T. Wheeler.
\newblock {Symmetric Solutions to the Gauss-Bonnet Extended Einstein
  Equations}.
\newblock {\em Nuclear Physics B}, 268(3):737--746, 1986.

\bibitem{whitt1988}
B.~Whitt.
\newblock {Spherically Symmetric Solutions of General Second Order Gravity}.
\newblock {\em Phys. Rev. D}, 38:3000, 1988.

\bibitem{Tibrewala2015}
R.~Tibrewala.
\newblock New second derivative theories of gravity for spherically symmetric
  spacetimes.
\newblock {\em Classical and Quantum Gravity}, 32:115007, 2015.

\bibitem{Louko1997}
Jorma Louko, Jonathan Simon, and Stephen Winters-Hilt.
\newblock {Hamiltonian Thermodynamics of a Lovelock Black Hole}.
\newblock {\em Physical Review D}, 55(6):3525--3535, 1997.

\bibitem{Taves2012}
Tim Taves, C.~Danielle Leonard, Gabor Kunstatter, and Robert~B. Mann.
\newblock {Hamiltonian Formulation of Scalar Field Collapse in
  Einstein--Gauss--Bonnet Gravity}.
\newblock {\em Classical and Quantum Gravity}, 29(1):015012, 2012.

\bibitem{Deser2005}
S.~Deser and J.~Franklin.
\newblock {Birkhoff for Lovelock Redux}.
\newblock {\em Classical and Quantum Gravity}, 22(16):L103--L106, 2005.

\bibitem{Teitelboim1987a}
Claudio Teitelboim and Jorge Zanelli.
\newblock {Dimensionally Continued Topological Gravitation Theory}.
\newblock {\em Classical and Quantum Gravity}, 4(4):125--129, 1987.

\bibitem{Hayward2006}
Sean~A. Hayward.
\newblock {Formation and Evaporation of Non-singular Black Holes}.
\newblock {\em Physical Review Letters}, 96:031103, 2006.

\bibitem{Frolov2014}
Valeri~P. Frolov.
\newblock {Information Loss Problem and a ‘Black Hole’ Model with a Closed
  Apparent Horizon}.
\newblock {\em Journal of High Energy Physics}, 2014(5):49, 2014.

\bibitem{Boulware1985}
D.~G. Boulware and S.~Deser.
\newblock {String-Generated Gravity Models}.
\newblock {\em Physical Review Letters}, 55(24):2656--2660, 1985.

\bibitem{Oliva2011}
J.~Oliva and R.~Sourya.
\newblock Birkhoff's theorem in higher derivative theories of gravity.
\newblock {\em Classical and Quantum Gravity}, 28:175007, 2011.

\bibitem{Myers2010}
R.~Myers and B.~Robinson.
\newblock Black holes in quasi-topological gravity.
\newblock {\em Journal of High Energy Physics}, 1008:067, 2010.

\end{thebibliography}
\bibliographystyle{unsrt} 


\end{document}